\documentclass{article}
\usepackage{graphicx}
\usepackage{amsmath,amssymb}
\usepackage{graphics,color,array,calc,rotating,epsfig,psfrag}
\usepackage{hyperref}
\numberwithin{equation}{section}
\usepackage{cite}
\usepackage{bm}
\usepackage{dcolumn}
\usepackage{float}
\usepackage{subfigure}
\usepackage{tikz-cd}
\usepackage{caption}
\usepackage[utf8]{inputenc}
\usepackage{tikz}
\usepackage{amsfonts}
\usepackage[mathscr]{eucal}
\usepackage{helvet}
\def\be{\begin{equation}} \def\ee{\end{equation}}
\def\bea{\begin{eqnarray}} \def\eea{\end{eqnarray}}

\newcommand{\RN}[1]{%
	\textup{\uppercase\expandafter{\romannumeral#1}}%
}
\newcommand{\sech}{\mathrm{sech}}
\newcommand{\csch}{\mathrm{csch}}
\begin{document}
	\baselineskip 18pt%
	\begin{titlepage}
		\vspace*{1mm}%
		\hfill%
		\vspace*{15mm}%
		\hfill
		\vbox{
			\halign{#\hfil         \cr
				%         hep-th/yymmnnn\cr
			%	IPM/P-2022/13  \cr
			} % end of \halign
		}  % end of \vbox
		\vspace*{20mm}
			\begin{center}
			{\Large {\bf \boldmath Generalized $T\bar{T}$-like flows  for scalar theories in two dimensions }}
		\end{center}
		\vspace*{5mm}
		\begin{center}
			{H. Babaei-Aghbolagh,$^{1,2}$ Song He,$^{1,2,3,4}$
				and Hao Ouyang$^{3}$
			}\\
			\vspace*{0.2cm}
			{\it
				$^{1}$Institute of Fundamental Physics and Quantum Technology, \\ Ningbo University, Ningbo, Zhejiang 315211, China \\
				$^{2}$School of Physical Science and Technology, 
				Ningbo University, Ningbo, 315211, China \\
				$^{3}$Center for Theoretical Physics and College of Physics, Jilin University, 
				Changchun 130012, China\\
				${}^4$Max Planck Institute for Gravitational Physics (Albert Einstein Institute),\\
				Am M\"uhlenberg 1, 14476 Golm, Germany\\
			}
			
			\vspace*{0.5cm}
			{E-mails: {\tt hosseinbabaei@nbu.edu.cn,  hesong@nbu.edu.cn, haoouyang@jlu.edu.cn
			}}
			\vspace{1cm}
		\end{center}
\begin{abstract}
  We demonstrate that the necessary condition for $SO(N) \times SO(N)$ duality invariance manifests as a partial differential equation in two-dimensional scalar theories. This condition, expressed as a partial differential equation, corresponds precisely to the integrability condition. We derive a general perturbation solution to this partial differential equation, which includes both a root $T\bar{T}$ flow equation and an irrelevant $T\bar{T}$-like flow equation. Additionally, we identify a general form for these flow equations that commute with each other. 
 Our results establish a general integrable theory characterized by theory-dependent coefficients at each order in the $\lambda$-expansion. This unified framework systematically classifies all integrable theories possessing two Lorentz-invariant variables ($P_1$, $P_2$) while accommodating arbitrary orders of the coupling constants ($\lambda$, $\gamma$). The theory provides a comprehensive classification scheme that encompasses both known and novel integrable systems within this class.

\end{abstract}
\end{titlepage}
%%%%%%%%%%%%%%%%%%%%%%%%%%%%%%%%%%%%%%%%%%%%%%%%%%%
\section{Introduction}\label{0}
%%%%%%%%%%%%%%%%%%%%%%%%%%%%%%%%%%%%%%%%%%%%%%%%%%%%
The profound connection between the Nambu-Goto theory in two dimensions and the Maxwell-Born-Infeld theory in four dimensions is well-documented \cite{Barbashov:1966frq, Barbashov:1966nvq, barb}. Notably, certain solutions of the Maxwell-Born-Infeld framework can also describe solutions of the Nambu-Goto theory when expressed in a static gauge with two transverse scalar fields. We consider a free scalar theory involving $N \geq 1$ scalar fields $\Phi^i$, where $i = 1, 2, \dots, N$. The Lagrangian for this theory is given by:
\begin{equation}\label{MS}
	{\cal L}_{\text{free}} = -\frac{1}{2} G_{ij} \partial_{\alpha}\Phi^i \partial^{\alpha}\Phi^j,
\end{equation}
where $G_{ij}$ is a symmetric tensor representing the moduli space metric of the scalar fields. Two-dimensional interacting scalar theories are constructed using two Lorentz-invariant variables, $P_1$ and $P_2$, with the Lagrangian expressed as ${\cal L}(\lambda, P_1, P_2)$, where $\lambda$ denotes the coupling constant. These variables are defined as:
\begin{equation}\label{P1P21}
	P_1 = G_{ij} \partial_{\alpha}\Phi^i \partial^{\alpha}\Phi^j, \quad 
	P_2 = G_{ik} G_{jl} \partial_{\alpha}\Phi^j \partial^{\alpha}\Phi^i \partial_{\beta}\Phi^l \partial^{\beta}\Phi^k.
\end{equation}
where $P_1$ and $P_2$ are the only two independent Lorentz invariant variables. All the scalar theories we consider are combinations of these two.

A two-dimensional free theory can undergo perturbative modification through a standard irrelevant $T \bar{T}$-deformation, as detailed in \cite{Smirnov:2016lqw, Cavaglia:2016oda}:
\begin{equation}\label{TTbar}
	O_{\lambda} = \frac{1}{8} \Big( {T_{\mu}}^{\mu} {T_{\nu}}^{\nu} - T_{\mu\nu}T^{\mu\nu} \Big).
\end{equation}
The deformed theory resulting from \eqref{TTbar} is a two-dimensional Born-Infeld theory:
\begin{eqnarray}\label{Msbi}
	{\cal L}_{BI-2D} = \frac{2}{\lambda} \bigg( 1 - \sqrt{1 + \frac{1}{2} \lambda P_1 + \frac{1}{8} \lambda^2 (P_1^2 - P_2)} \bigg),
\end{eqnarray}
where $\lambda$ is the dimensional deformation parameter.

In four dimensions, marginal root deformations \cite{Babaei-Aghbolagh:2022MoxMax} represent a distinct class of deformations governed by a dimensionless parameter $\gamma$. For example,  ModMax theory \cite{Bandos:2020jsw} is derived from Maxwell theory through root flow perturbation \cite{Babaei-Aghbolagh:2022MoxMax}. The operator generating this marginal root-deformed theory in two dimensions is:
\begin{eqnarray}\label{Margi}
	{\cal R}_{\gamma} = \frac{1}{\sqrt{2}} \sqrt{T_{\mu\nu}T^{\mu\nu} - \frac{1}{2} {T_{\mu}}^{\mu} {T_{\nu}}^{\nu}},
\end{eqnarray}
known as the root $T \bar{T}$ deformation operator.

After presenting the ModMax theory in four dimensions in Ref.~\cite{Bandos:2020jsw} and the flow equation based on the $\gamma$ coupling in Ref.~\cite{Babaei-Aghbolagh:2022MoxMax}, the initial version of Ref.~\cite{Conti:2022egv} introduced a two-dimensional ModMax theory. Subsequently, three papers appeared in arXiv : the second version of Ref.~\cite{Conti:2022egv} and Ref.~\cite{ Ferko:2206jsw} were appeared on the same day, followed five days later by Ref.~\cite{Babaei-Aghbolagh:2022kjj}, which presented the $\gamma$-flow equation for the two-dimensional ModMax theory in the form of operator~\eqref{Margi}.
Studies highlight connections between $T \bar{T}$-like flows and gravity theories in various dimensions, particularly in models coupled to gravity, using the vielbein formalism for deformed field theories \cite{Babaei2024c, Tsolakidis2024cw, Ferko2024cw, Morone:2024egv, Allameh:2021moy}. For recent advances on root-type $T \bar{T}$ deformation, see \cite{Hou:2022csf, Garcia:2022wad, Tempo:2022ndz, Bagchi:2022tcz, Bagchi:2024cz, Ferko:2023ruw, Ferko:2023sps, Ebert:2023tih, Rodriguez:2021tcz, Ferko:2023ozb, Ferko:2023iha, he2024irrelevan, ferko2024interacting, Tian2024c}.

The operator in \eqref{Margi} transforms the perturbative structure of the free action into a two-dimensional Scalar Modified Maxwell (SMM) theory \cite{Babaei-Aghbolagh:2022kjj}, described by:
\begin{equation}\label{SMMdw}
	{\cal L}_{SMM}(\gamma) = -\frac{1}{2} \Big( \cosh(\gamma) P_1 + \sinh(\gamma) \sqrt{-P_1^2 + 2P_2} \Big).
\end{equation}

Applying the irrelevant and marginal operators \eqref{TTbar} and \eqref{Margi} to the free theory \eqref{MS} yields a Generalized Scalar ModMax (GSMM) theory, defined as:
\begin{eqnarray}\label{GSMMdw}
	{\cal L}_{GSMM} = \frac{2}{\lambda} \bigg( 1 - \sqrt{1 + \frac{1}{2}\lambda \bigg( \cosh(\gamma) P_1 + \sinh(\gamma) \sqrt{-P_1^2 + 2P_2} \bigg) + \frac{1}{8} \lambda^2 (P_1^2 - P_2)} \bigg).
\end{eqnarray}
The GSMM theory satisfies the following flow equations with respect to $\gamma$ and $\lambda$:
\begin{align}\label{lM2}
	\boxed{
		\frac{\partial {\cal L}_{GSMM}}{\partial \lambda} = \frac{1}{8} \big( T_{\mu\nu}T^{\mu\nu} - {T_{\mu}}^{\mu} {T_{\nu}}^{\nu} \big), \quad
		\frac{\partial {\cal L}_{GSMM}}{\partial \gamma} = \frac{1}{\sqrt{2}} \sqrt{T_{\mu\nu}T^{\mu\nu} - \frac{1}{2} {T_{\mu}}^{\mu} {T_{\nu}}^{\nu}}
	}.
\end{align}
Refs. \cite{Babaei-Aghbolagh:2022kjj,Ferko:2206jsw} demonstrates that operators \eqref{TTbar} and \eqref{Margi} commute, as illustrated in Fig.~\eqref{fig1}.

\begin{center}
	\begin{tikzcd}[row sep=huge,column sep=huge]
		{\cal L}_{free} \arrow[r, blue, "{O}_{\lambda}" blue] \arrow[d,red,"{\cal R}_{\gamma}" red]
		&|[blue]| {\cal L}_{BI-2D} \arrow[d,red, "{\cal R}_{\gamma}" red] \\
		|[red]|{\cal L}_{SMM} \arrow[r, blue, "{O}_{\lambda}" blue]
		&|[red!50!blue]|  {\cal L}_{GSMM}
	\end{tikzcd}
	\captionof{figure}{Deformations of the  multi-scalar theories under ${O}_{\lambda}$ and ${\cal R}_{\gamma}$.}\label{fig1}
\end{center}

This approach involves preserving a specific symmetry in a theory by ensuring the Lagrangian satisfies a PDE with respect to the theory's parameters. For example, the SO(2) symmetry in four-dimensional electrodynamics leads to the following PDE \cite{Bialynicki-Birula:1981, Bialynicki-Birula:1992rcm, Gaillard:1981rj, Gaillard:1997rt, Gibbons:1995ap}: 
\begin{eqnarray}
	\label{PDEM} 
	\big( (\partial_t \mathcal{L})^2 - (\partial_z \mathcal{L})^2 - 1 \big) z - \big( 2 (\partial_z \mathcal{L}) (\partial_t \mathcal{L}) \big) t = 0,
\end{eqnarray}
where $\mathcal{L}_t = \frac{\partial \mathcal{L}}{\partial t}$ and $\mathcal{L}_z = \frac{\partial \mathcal{L}}{\partial z}$, with $t = \frac{1}{4} F_{\mu\nu} F^{\mu\nu}$ and $z = \frac{1}{4} F_{\mu\nu} \tilde{F}^{\mu\nu}$ being two Lorentz-invariant variables. 

The solutions to this equation are theories exhibiting SO(2) symmetry. For instance, the nonlinear electrodynamic theories of Born-Infeld \cite{Born:1933lls, Born:1934gh}, ModMax \cite{Bandos:2020jsw}, and General ModMax \cite{Bandos:2020hgy, Sorokin:2021tge} are all solutions to the PDE~\eqref{PDEM}. The relationship between electromagnetic duality and $T \bar{T}$-like deformations at the effective action level has been explored in \cite{Conti:2018jho, Kuzenko:2000uh, BabaeiVelni:2016qea, BabaeiVelni:2019ptj, Babaei-Aghbolagh:2013hia, Babaei-Aghbolagh:2020kjg, Aghbolagh2210, Kuzenko2024c, Christian2024cw}. 
Additionally, innovative methods for solving this equation can be found in \cite{IZ1, Ivanov:2004jv, INZ, Mkrtchyan:2205uvc, Mkrtchyan:2019opf, russo2024causal, russo2024dual, Babaei-Aghbolagh:2024jho}. The approach has also been extended to the Carrollian framework \cite{Chen2024c}, where a Carrollian electrodynamics theory with both conformal invariance and SO(2) duality is explored by solving a PDE.
In \cite{KMcA,KMcA23}, sigma models were introduced that are invariant under U(1) "duality rotations," which exchange the dynamical variables and their equations of motion. The Lagrangians of these sigma models obey a partial differential equation analogous to the self-duality equation followed by U(1) duality invariant models in nonlinear electrodynamics.

T-duality, a central concept in perturbative string theory, emerges when the theory is compactified on a torus \cite{Giveon:1994fu}. This compactification reveals that the spectrum of the free string on a torus remains invariant under $O(N, N)$ transformations \cite{Sen:1991zi, Hohm:2014sxa}. This symmetry extends beyond the free string spectrum to the full bosonic string theory, where compactification on a torus $T^N$ preserves invariance under $O(N, N)$ transformations \cite{Garousi:2019mca}. More notably, after integrating the massive modes, T-duality manifests as symmetry in the effective actions \cite{Garousi:2017fbe}. It has been demonstrated that the dimensional reduction of the classical effective actions of bosonic string theory at each order in $\alpha'$ is invariant under $O(N, N)$ transformations \cite{Garousi:2017fbe, Garousi:2020gio}. This insight has led to the development of sigma models, where the worldsheet string theory remains invariant under $O(N, N)$ transformations in two dimensions. 
When the target space metric is $G_{ij} = \delta_{ij}$, and the Kalb-Ramond field vanishes, the symmetry group $O(N, N)$ reduces to $SO(N) \times SO(N)$. In Section~\eqref{Appen}, we show that the necessary and sufficient condition for invariance under $SO(N) \times SO(N)$ transformations is given by the following PDE, which involves the two Lorentz-invariant variables $P_1$ and $P_2$: 
\begin{eqnarray}
	\label{lagrangeNGZ2D} 
	8 (P_1^2 - P_2) \left( \frac{\partial \mathcal{L}}{\partial P_2} \right)^2 + 8 P_1 \frac{\partial \mathcal{L}}{\partial P_2} \frac{\partial \mathcal{L}}{\partial P_1} + 4 \left( \frac{\partial \mathcal{L}}{\partial P_1} \right)^2 - 1 = 0.
\end{eqnarray}
This PDE provides a unified solution framework for both free and deformed theories, such as~\eqref{MS},~\eqref{Msbi},~\eqref{SMMdw}, and~\eqref{GSMMdw} (see Refs. \cite{Borsato:2022tmu, Ebert:2024zwv}), which are classified together due to their adherence to the condition in~\eqref{lagrangeNGZ2D}.
Refs. \cite{Borsato:2022tmu, Ebert:2024zwv} demonstrate that the model achieves classical integrability when the Lagrangian satisfies Eq.~\eqref{lagrangeNGZ2D}. Therefore, the criterion for classical integrability of the model, as outlined in condition~\eqref{lagrangeNGZ2D}, coincides with the PDE that a Lagrangian must obey in a four-dimensional theory of duality-invariant electrodynamics.
The integrability condition for all integrable theories composed of two variables $P_1$ and $P_2$ manifests as the differential equation presented in Eq.~\eqref{lagrangeNGZ2D} to all orders in $\lambda$. Notably, both ModMax theory and the generalized Born-Infeld theory emerge as particular cases within this broader class of theories.
	In this work, we introduce a novel systematic approach for generating all such integrable theories by solving the integrability condition given in Eq.~\eqref{lagrangeNGZ2D}. Our methodology possesses the following key features:
	\begin{itemize}
		\item In the absence of computational constraints, this approach yields exact, integrable solutions to all orders in $\lambda$ through a deformation procedure
		\item  The perturbative solution can provide a framework for understanding integrable theories in this class.
        
		\item We explicitly derive a family of non-trivial solutions through $\mathcal{O}(\lambda^n)$, where $n$ is an arbitrary integer
\end{itemize}

This paper aims to identify two categories of general solutions to the PDE~\eqref{lagrangeNGZ2D}, corresponding to distinct marginal and irrelevant $T \bar{T}$-like deformations. We will delineate the general structure of these deformations using marginal and irrelevant flow equations. Specifically, we classify theories of non-linear sigma models dependent on two Lorentzian variables, $P_1$ and $P_2$, which feature two types of coupling parameters: a dimensionless coupling $\gamma$ and a dimensional coupling $\lambda$. We will derive expressions for the $\gamma$ and $\lambda$ couplings from the root-type and irrelevant transformations of $T \bar{T}$.

The organization of this paper is as follows: In Section~\eqref{22}, we show that the invariance condition under $SO(N) \times SO(N)$ is governed by the PDE in~\eqref{lagrangeNGZ2D}, which is identical to the equation derived from the integrability condition in Equation 3.32 of Ref. \cite{Borsato:2022tmu}. In Section~\eqref{2.23}, we solve the PDE in~\eqref{lagrangeNGZ2D} using perturbation theory. This method considers all possible solutions in integer powers of $P_1$ and $P_2$ at each order of $\lambda$, determining the associated constants. We demonstrate that an irrelevant flow equation exists for this solution, constructed from integer powers of the $T_{\mu \nu} T^{\mu \nu}$ and $T^{\mu}{}_{\mu} T^{\nu}{}_{\nu}$ structures, and is commutative with root deformation. In Section~\eqref{Sec3}, we transform the PDE~\eqref{lagrangeNGZ2D} into a new differential form using the variables $P_1$ and $\mathcal{P}$, where $\mathcal{P} := \sqrt{- P_1^2 + 2 P_2}$. This transformation allows us to derive a general perturbation solution and to determine how the solution coefficients depend on the $\gamma$ coupling, yielding a root flow equation. In Section~\eqref{0409}, we identify the general form of the irrelevant flow equations for these theories, which commute with the root deformation. Finally, in Section~\eqref{04}, we summarize our results and discuss future directions.
%%%%%%%%%%%%%%%%%%%%%%%%%%%%%%%%%%%%%%%%%%%%%%%%%%%%%%%%%%%%%%%%%%%%%%%%%%%%%%%%%%%%%%%%%%%%%%%%%%%%%%%%%%%%%%%%%%%%%%%%%%%%%%%%%%%%%%%%%%%%%%%%%%%%%%%%%%%%%%%%%%%%%%%%%%%%%%%%%%%%%%%%%%%%%%%%%%%%%%%%%%%%%%%%%%
\section{ Integrability condition vs.  $SO(N) \times SO(N)$ duality invariance  }\label{22}
%%%%%%%%%%%%%%%%%%%%%%%%%%%%%%%%%%%%%%%%%%%%%%%%%%%%%%%%%%%%%%%%%%%%%%%%%%%%%%%%%%%%%%%%%

The integrability condition in Eq.~\eqref{lagrangeNGZ2D} has recently been identified as a necessary and sufficient condition for $O(N, N)$ duality in two-dimensional scalar theories. These theories play a central role in integrable systems and are intimately connected with the integrable models discussed in this section. We explore how the invariance under $SO(N) \times SO(N)$ transformations emerges as a crucial feature for maintaining integrability and investigate the relationship between these two concepts in the context of the scalar theory.

%%%%%%%%%%%%%%%%%%%%%%%%%%%%%%%%%%%%%%%%%%%%%%%%%%%
\subsection{Integrability condition as a PDE}\label{1.}
%%%%%%%%%%%%%%%%%%%%%%%%%%%%%%%%%%%%%%%%%%%%%%%%%%%%
Refs. \cite{Borsato:2022tmu, Ebert:2024zwv} present a detailed stress tensor analysis within a generic interacting chiral boson theory. The study carefully examines the classical flow properties governed by the stress tensor $T_{\mu\nu}$, focusing on maintaining Lorentz invariance. In Ref. \cite{Ebert:2024zwv}, the authors derive a Lorentz-invariant condition for the Lagrangian of the principal chiral model (PCM). \footnote{The Lorentz condition for chiral scalar theory is given in Eq. (2.35) of Ref. \cite{Ebert:2024zwv} as 
	$  \bigl(\frac{\partial \mathcal{L}}{\partial S}\bigr)^2 + 2 \frac{S}{P} \frac{\partial \mathcal{L}}{\partial S} \frac{\partial \mathcal{L}}{\partial P} + \bigl(\frac{\partial \mathcal{L}}{\partial P}\bigr)^2 - 1 = 0 $. This is reformulated in Eq.~\eqref{lagrangeNGZ2D} by substituting $S \Longleftrightarrow - \frac{1}{2} P_1$ and $P^2 \Longleftrightarrow \frac{P_1^2 - P_2}{2} $. The variables $S$ and $P$ are defined as $S = \frac{1}{2} \left( \phi^{\prime j} \phi^{\prime j} + \bar{\phi}^{\prime \bar{j}} \bar{\phi}^{\prime \bar{j}} \right)$ and $P = \frac{1}{2} \left( \phi^{\prime j} \phi^{\prime j} - \bar{\phi}^{\prime \bar{j}} \bar{\phi}^{\prime \bar{j}} \right)$.} This condition is crucial for ensuring that the interaction function accurately reflects a Lorentz-invariant theory.

Studying integrable structures in string theory has further fueled the search for transformations that preserve integrality in two-dimensional integrable quantum field theories. In particular, identifying integrable deformations in 2D sigma models is relevant for string theory applications. We focus on classical field theories defined on a flat, two-dimensional spacetime manifold, denoted as $ \Sigma $, which we occasionally refer to as the worldsheet. Coordinates $ \sigma^{\alpha} = (\sigma, \tau) $ are chosen on $ \Sigma $. Our primary interest lies in two-dimensional sigma models where the target space $ G $ is a Lie group, and its Lie algebra is denoted as $ \mathfrak{g} $. The fundamental field $ g(\sigma, \tau) $ maps the worldsheet $ \Sigma $ into the Lie group $ G $. From this field $ g $, we can construct two important quantities: the left- and right-invariant Maurer-Cartan forms, defined as follows:
\begin{equation}
	\label{P1P21}
	j=g^{-1} dg , \quad \quad \quad  \tilde{j}= -(dg) g^{-1}\,
.
\end{equation}
Both $j$ and $\tilde{j}$ satisfy the flatness condition, expressed using light-cone coordinates\footnote{We take  for any vector $A^\mu$, $A^\pm = \tfrac{1}{2}(A^0\pm A^1)$.}:
\begin{equation}
	\label{jj}
	\partial_+ j_- -\partial_- j_+ + \left[j_+,\,j_-\right]=0=\partial_+ \tilde{j}_- -\partial_- \tilde{j}_+ + \left[\tilde{j}_+,\,\tilde{j}_-\right] \,.
\end{equation}
One of the simplest examples of such a unified sigma model is the Principal Chiral Model (PCM). The Lagrangian of the principal chiral model is expressed in terms of $j$ or  $\tilde{j}$:
\begin{equation}
	\label{eq:Lpcm}
	\mathcal{L}_{PCM} = \frac{1}{2}g^{\mu\nu}\text{tr}\left[j_\mu\,j_\nu\right]=-\frac{1}{2}\text{tr}\left[j_+\,j_-\right]\,,
\end{equation}
which can be equivalently written as
\begin{equation}
	\mathcal{L}_{PCM} = \frac{1}{2}g^{\mu\nu}\text{tr}\left[\tilde{j}_\mu\,\tilde{j}_\nu\right]=-\frac{1}{2}\text{tr}\left[\tilde{j}_+\,\tilde{j}_-\right]\,.
\end{equation}
% One of the simplest examples of such a unified sigma model is the Principal Chiral Model (PCM), whose Lagrangian is as follows:
%\begin{eqnarray}
%	\label{LPCM} 
%	\mathcal{L}_{PCM} =\frac{ 1}{2} G_{ij} \partial_{\alpha}\Phi^{i} \partial^{\alpha}\Phi^{j}
%\end{eqnarray}
Any Lagrangian that depends on \( j_{\pm} \) only through the combinations of $\,\, \text{tr}[j_+ j_-] \,\,$ and $\text{tr}(j_+ j_+)\,\, \text{tr}[j_- j_-] $ can be expressed, after a change of variables, as a function $\mathcal{L}(P_1, P_2)$ of the two variables:
\begin{equation}
	\label{P1P2}
	P_1 := -\text{tr}[j_+ j_-]\,,\qquad
	P_2 := \frac{1}{2}\left(\text{tr}[j_+j_+]\text{tr}[j_-j_-]+\left(\text{tr}[j_+ j_-]\right)^2\right) \, 
.
\end{equation}
The equation of motion for any Lagrangian $ \mathcal{L}(P_1, P_2)$ can be expressed as $ \partial_\alpha \mathfrak{J}^\alpha = 0 $, where $ \mathfrak{J}^\alpha $ is the Noether current associated with the symmetry under right-multiplication of $ g$ by a general group element. The Noether current for any such Lagrangian $ \mathcal{L}(P_1, P_2) $ is:
\begin{equation}
	\label{eq:deformedPCMeomgeneric}
	\mathfrak{J}_\mu =2\frac{\partial \mathcal{L}(P_1, P_2)}{\partial P_1} j_\mu +
	4\frac{\partial \mathcal{L}(P_1, P_2)}{\partial P_2} g^{\nu\rho}\text{tr}[j_\mu j_\nu]\,j_\rho\, .
\end{equation}
The equations of motion for the deformed theories, characterized by the couplings $\gamma$ and $\lambda$ and dependent on the independent variables $P_1$ and $P_2$, are equivalent to the flatness condition of the Lax connection 
\begin{equation}
	\mathfrak{L}_{\pm}^{(\lambda,\gamma)} = \frac{j_\pm \pm z\,\mathfrak{J}_{\pm}}{1-z^2}\,
.
\end{equation}
The flatness condition for $\mathfrak{L}^{(\lambda,\gamma)}$ is given by:
\begin{equation}
	\label{flatness}
	0=\partial_{+}\mathfrak{L}^{(\lambda,\gamma)}_{-} - \partial_{-}\mathfrak{L}^{(\lambda,\gamma)}_{+} +[\mathfrak{L}^{(\lambda,\gamma)}_+,\mathfrak{L}^{(\lambda,\gamma)}_-]\,.
\end{equation}The flatness condition in equation~\eqref{flatness} is met when the current $\mathfrak{J}_{\alpha}$ satisfies the property $ \big[\mathfrak{J}_+,j_-\big]=\big[j_+,\mathfrak{J}_-\big]  $.As shown in Ref.  \cite{Borsato:2022tmu}, this equality holds if and only if the Lagrangian $\mathcal{L}(P_1, P_2) $ is satisfied the PDE in~\eqref{lagrangeNGZ2D}.
%%%%%%%%%%%%%%%%%%%%%%%%%%%%%%%%%%%%%%%%%%%%%%%%%%%%%%%%%%%%%%%%%%%%%%%%%%%%
\subsection{$SO(N)\times SO(N)$ duality invariance}\label{Appen}
%%%%%%%%%%%%%%%%%%%%%%%%%%%%%%%%%%%%%%%%%%%%%%%%%%%%%%%%%%%%%%%%%
It is well known the worldsheet string theory remains invariant under $O(N, N)$ transformations in two dimensions. This invariance has led to the development of sigma models \cite{Sen:1991zi, Hohm:2014sxa}.
Two-dimensional non-linear sigma models admit $O(N, N)$ duality transformations (a review can be found in \cite{Giveon:1994fu}), which includes T-duality as a special case. When the target space metric $G_{ij}=\delta_{ij}$ and 
the Kalb-Ramond field is zero, a non-linear sigma model is invariant under a $SO(N)\times SO(N)$ subgroup duality transformation. 
In this subsection, we show that the condition~\eqref{lagrangeNGZ2D} leads to the $SO(N)\times SO(N)$ duality invariance for scalar theories ${\cal L}(P_1,P_2)$ with $G_{ij}=\delta_{ij}$. We denote
\begin{eqnarray}    
	F_{\mu}^{~i}&=&\partial_\mu \phi^i,\\
	H_{\mu}^{~i}&=&-\epsilon_{\mu\nu}\frac{\partial \mathcal{L}}{\partial \partial_\nu \phi^i}, \label{defH}
\end{eqnarray}
where the convention of Levi-Civita symbol is $\epsilon_{01}=1$. In the following, we will use matrix notation. For instance, we denote $\eta^{\mu\nu}\partial_\mu \phi^i \partial_\nu \phi^i=\mathrm{Tr}(F^t \eta^{-1} F)$ and~\eqref{defH} can be written as
\begin{equation}
	d  \mathcal{L}= \mathrm{Tr}(H^t \epsilon^{-1} d F). \label{defH2}
\end{equation}
The equations of motion of $F$ and $H$ are
\begin{align}    
	\partial_\nu (\epsilon^{-1})^{\nu \mu}F_{\mu}^{~i}&=0,\\
	\partial_\nu (\epsilon^{-1})^{\nu \mu} H_{\mu}^{~i}&=0.
\end{align}
They are invariant under the $SO(N)\times SO(N)$ duality rotation:
\begin{equation}\label{dualitytran}
	F+H \rightarrow (F+H) O_1 ,~~~
	F-H\rightarrow (F-H) O_2,  
\end{equation}  
where $O_1$ and $O_2$ are two $N\times N$ special orthogonal matrices. The diagonal subgroup defined by  $O_1=O_2=O$ is the usual $SO(N)$ symmetry $\phi^j\rightarrow \phi^i O_i^{~j}$. The genuine duality can be obtained as the coset modulo the diagonal subgroup. When $N=2$, this duality transformation is related to the duality of four-dimensional electrodynamics via dimensional reduction. In the treatment of duality, $F$ and $H$  are not seen as functionals of $\phi$, and  are constrained by~\eqref{defH} or equivalently~\eqref{defH2}, where $\mathcal{L}$ is regarded a function of $F$.

The duality invariance requires that~\eqref{defH2} is invariant under the transformations~\eqref{dualitytran}. We now show that~\eqref{lagrangeNGZ2D} leads to duality invariance.
For an arbitrary $\mathcal{L}$, $H$ can be computed as
\begin{equation}
	H=-2 \frac{\partial \mathcal{L}}{\partial P_1} \epsilon \eta^{-1} F-4 \frac{\partial \mathcal{L}}{\partial P_2} \epsilon \eta^{-1} F F^t \eta^{-1} F.
\end{equation}
For any antisymmetric matrix $b$ with internal indices of $SO(N)$, we have
\begin{equation}
	\mathrm{Tr}(H^t \epsilon^{-1} H b)
	=- \mathrm{Tr}(F^t \epsilon^{-1} F b) \left[ 8 (P_1^2 -  P_2) \bigl(\frac{\partial \mathcal{L}}{\partial P_2}\bigr)^2 + 8 P_1 \frac{\partial \mathcal{L}}{\partial P_2} \frac{\partial \mathcal{L}}{\partial P_1} + 4 \bigl(\frac{\partial \mathcal{L}}{\partial P_1}\bigr)^2 \right].
\end{equation}
To derive the formula, we use the observation that $F b F^t$ is a $2\times 2$ antisymmetric matrix, and therefore, the trace can be factorized. When the condition~\eqref{lagrangeNGZ2D} is satisfied, we find
\begin{equation}
	\mathrm{Tr}(H^t \epsilon^{-1} H b)   +\mathrm{Tr}(F^t \epsilon^{-1} F b) =0.
\end{equation}
Another useful identity is
\begin{equation}
	\mathrm{Tr}(H^t \epsilon^{-1} F b)=\mathrm{Tr}(F^t \epsilon^{-1} H b) =0.
\end{equation}

The variances of the infinitesimal duality rotation are
\begin{equation}
	\delta F=F a+ H b,~~~\delta H=F b+H a,
\end{equation}
where $a$ and $b$ are two infinitesimal antisymmetric matrices.
Then we get
\begin{equation}
	\mathcal{L}(F+\delta F)-\mathcal{L}(F)
	=\mathrm{Tr}(H^t \epsilon^{-1} \delta F) =\frac{1}{2}\mathrm{Tr}(H^t \epsilon^{-1} H b)   -\frac{1}{2}\mathrm{Tr}(F^t \epsilon^{-1} F b) .
\end{equation}
Taking derivative, we find
\begin{eqnarray}
	d\left(\mathcal{L}(F+\delta F)-\mathcal{L}(F)\right)
	&=&\mathrm{Tr}(H^t \epsilon^{-1} dH b)   -\mathrm{Tr}(F^t \epsilon^{-1} dF b) \nonumber\\
	&=&\mathrm{Tr}(H^t \epsilon^{-1} d(\delta F)   +\mathrm{Tr}(\delta H^t \epsilon^{-1} dF ),\\
	\Rightarrow 
	d\mathcal{L}(F+\delta F)
	&=&
	\mathrm{Tr}((H+\delta H)^t \epsilon^{-1} d (F+\delta F)).
\end{eqnarray}
Therefore, the constraint~\eqref{defH2} is invariant under the duality rotation, and the condition~\eqref{lagrangeNGZ2D} leads to the $SO(N)\times SO(N)$ duality invariance.

In \cite{Bielli:2024khq}, it was proved that the T-duality transformation commutes with stress-energy tensor transformations using an auxiliary
field formulation \cite{Ferko2024cw, Bielli:2024ach}. This result can be easily extended to the $O(N, N)$ duality transformations.
Consequently, stress-energy tensor transformations preserve the $SO(N)\times SO(N)$ duality invariance, which explains that stress-energy tensor transformations can realize perturbative solutions of~\eqref{lagrangeNGZ2D}.

%%%%%%%%%%%%%%%%%%%%%%%%%%%%%%%%%%%%%%%%%%%%%%%%%%%%%%%%%%%%%%%%%%%%%%%%%%%%%%%%%%%%%%%%%%%%%%%%%%%%%%%%%%%%%%%%%%%%%%%%%%
\section{General irrelevant  $T \bar{T}$-like deformations for perturbative solutions }\label{2.23}
%%%%%%%%%%%%%%%%%%%%%%%%%%%%%%%%%%%%%%%%%%%%%%%%%%%%%%%%%%%%%%%%%%%%%%%%%%%%%%%%%%%%%%%%%%%%%%%%%%%%%%%%%%%%%%%%%%%%%%%%%%%%%%%%%%%%%%%%%%%%%%%%%%%%%%%%%%%%%%%%%%%%%%%%%%%%%%%%%%%%%%%%%%%%%%%%%%%%%%%%%%%%%
We investigate the perturbative solutions of the PDE~\eqref{lagrangeNGZ2D}, described by general irrelevant  $T \bar{T}$-like deformations. This type has a dimensionful $\lambda$ coupling, and we take it as a function of the integers power of $P_1$ and $P_2$ in the form of $\mathcal{L}(\lambda^n)=\mathcal{K}(P_1^N, P_2^M,\lambda^{N+2M-1})$. The Lagrangian can be expressed up to the $\lambda^7$ order, incorporating arbitrary coefficients denoted by $a_n$, in the following manner:
\begin{eqnarray}
	\label{lagrangeGeneral}
	\mathcal{L}(\lambda^n)&=&a_{1}{} P_{1}{} + \lambda \bigl(a_{2}{} {P_{1}}^2 + a_{3}{} P_{2}{}\bigr) + \lambda^2 \bigl(a_{4}{} {P_{1}}^3 + a_{5}{} P_{1}{} P_{2}{}\bigr) + \lambda^3 \bigl(a_{6}{} {P_{1}}^4 + a_{7}{} {P_{1}}^2 P_{2}{} + a_{8}{} {P_{2}}^2\bigr) \nonumber\\
	&& + \lambda^4 \bigl(a_{9}{} {P_{1}}^5 + a_{10}{} {P_{1}}^3 P_{2}{} + a_{11}{} P_{1}{} {P_{2}}^2\bigr)  \nonumber\\
	&&+ \lambda^5 \bigl(a_{12}{} {P_{1}}^6 + a_{13}{} {P_{1}}^4 P_{2}{} + a_{14}{} {P_{1}}^2 {P_{2}}^2 + a_{15}{} {P_{2}}^3\bigr) \nonumber\\
	&& + \lambda^6 \bigl(a_{16}{} {P_{1}}^7 + a_{17}{} {P_{1}}^5 P_{2}{} + a_{18}{} {P_{1}}^3 {P_{2}}^2 + a_{19}{} P_{1}{} {P_{2}}^3\bigr) \nonumber\\
	&& + \lambda^7 \bigl(a_{20}{} {P_{1}}^8 + a_{21}{} {P_{1}}^6 P_{2}{} + a_{22}{} {P_{1}}^4 {P_{2}}^2 + a_{23}{} {P_{1}}^2 {P_{2}}^3 + a_{24}{} {P_{2}}^4\bigr)\,\nonumber\\
	&& + \lambda^8 \bigl(a_{25}{} {P_{1}}^9 + a_{26}{} {P_{1}}^7 P_{2}{} + a_{27}{} {P_{1}}^5 {P_{2}}^2 + a_{28}{} {P_{1}}^3{P_{2}}^3 + a_{29}{} P_{1}{} {P_{2}}^4\bigr)\,.
\end{eqnarray}
This section aims to solve the PDE sequentially~\eqref{lagrangeNGZ2D}, utilizing the Lagrangian of order $\lambda^7$.
This process will clarify the relationship between the coefficients $a_n$ at each order of $\lambda$, ultimately leading to the following correlation among the coefficients $a_n$. This correlation is detailed in Appendix~\eqref{Appenn}.
Upon applying conditions~\eqref{an} to the Lagrangian~\eqref{lagrangeGeneral}, a solution to equation~\eqref{lagrangeNGZ2D} emerges, involving several constants $a_2$, $a_7$, $a_{13}$ and $a_{21}$ which remain undetermined. At the order of $ \lambda^n$, $ \frac{n}{2} $ coefficients are not fixed. This general Lagrangian is obtained as follows:
\begin{eqnarray}
	\label{Lpp}
	{\cal L}(\lambda)&=&- \tfrac{1}{2} P_{1}{} + \lambda a_{2}{} \bigl({P_{1}}^2 - 2 P_{2}{}\bigr) + 4 \lambda^2 {a_{2}}^2 P_{1}{} \bigl({P_{1}}^2 - 2 P_{2}{}\bigr)  \\
	&&+ \lambda^3 \Bigl(- \tfrac{1}{4} a_{7}{} \bigl({P_{1}}^2 - 2 P_{2}{}\bigr)^2 + 8 {a_{2}}^3 \bigl({P_{1}}^4 - 4 {P_{2}}^2\bigr)\Bigr) \nonumber\\
	&& + \lambda^4 \Bigl(-4 a_{2}{} a_{7}{} P_{1}{} \bigl({P_{1}}^2 - 2 P_{2}{}\bigr)^2 - 16 {a_{2}}^4 P_{1}{} \bigl(3 {P_{1}}^4 - 20 {P_{1}}^2 P_{2}{} + 28 {P_{2}}^2\bigr)\Bigr)  \nonumber\\
	&&+ \lambda^5 \Bigl(- \tfrac{1}{6} a_{13}{} \bigl({P_{1}}^2 - 2 P_{2}{}\bigr)^3 -  \tfrac{40}{3} {a_{2}}^2 a_{7}{} \bigl({P_{1}}^2 - 2 P_{2}{}\bigr)^2 \bigl({P_{1}}^2 + 4 P_{2}{}\bigr) \nonumber\\
	&& -  \tfrac{512}{3} {a_{2}}^5 \bigl({P_{1}}^6 - 18 {P_{1}}^2 {P_{2}}^2 + 28 {P_{2}}^3\bigr)\Bigr) \nonumber\\
	&& + \lambda^6 \Bigl(32 {a_{2}}^3 a_{7}{} P_{1}{} \bigl(11 {P_{1}}^2 - 42 P_{2}{}\bigr) \bigl({P_{1}}^2 - 2 P_{2}{}\bigr)^2 + {a_{7}}^2 P_{1}{} \bigl({P_{1}}^2 - 2 P_{2}{}\bigr)^3  \nonumber\\
	&&- 4 a_{2}{} a_{13}{} P_{1}{} \bigl({P_{1}}^2 - 2 P_{2}{}\bigr)^3 + 128 {a_{2}}^6 \bigl(61 {P_{1}}^7 - 454 {P_{1}}^5 P_{2}{} + 1116 {P_{1}}^3 {P_{2}}^2 - 904 P_{1}{} {P_{2}}^3\bigr)\Bigr) \nonumber\\
	&& + \lambda^7 \Bigl(- \tfrac{1}{8} a_{21}{} \bigl({P_{1}}^2 - 2 P_{2}{}\bigr)^4 + 7 a_{2}{} {a_{7}}^2 \bigl({P_{1}}^2 - 2 P_{2}{}\bigr)^3 \bigl({P_{1}}^2 + 6 P_{2}{}\bigr)  \nonumber\\
	&&- 14 {a_{2}}^2 a_{13}{} \bigl({P_{1}}^2 - 2 P_{2}{}\bigr)^3 \bigl({P_{1}}^2 + 6 P_{2}{}\bigr) + 448 {a_{2}}^4 a_{7}{} \bigl({P_{1}}^2 - 2 P_{2}{}\bigr)^2 \bigl(3 {P_{1}}^4 + 12 {P_{1}}^2 P_{2}{} - 56 {P_{2}}^2\bigr) \nonumber\\
	&& + 4096 {a_{2}}^7 \bigl(7 {P_{1}}^8 - 206 {P_{1}}^4 {P_{2}}^2 + 608 {P_{1}}^2 {P_{2}}^3 - 504 {P_{2}}^4\bigr)\Bigr) \nonumber\\
	&& + \lambda^8 \Bigl(\tfrac{32}{3} {a_{2}}^3 a_{13}{} P_{1}{} \bigl(73 {P_{1}}^2 - 258 P_{2}{}\bigr) \bigl({P_{1}}^2 - 2 P_{2}{}\bigr)^3 - 8 {a_{2}}^2 {a_{7}}^2 P_{1}{} \bigl(65 {P_{1}}^2 - 242 P_{2}{}\bigr) \bigl({P_{1}}^2 - 2 P_{2}{}\bigr)^3 \nonumber\\
	&& + 2 a_{7}{} a_{13}{} P_{1}{} \bigl({P_{1}}^2 - 2 P_{2}{}\bigr)^4 - 4 a_{2}{} a_{21}{} P_{1}{} \bigl({P_{1}}^2 - 2 P_{2}{}\bigr)^4 \nonumber\\
	&& -  \tfrac{128}{3} {a_{2}}^5 a_{7}{} P_{1}{} \bigl({P_{1}}^2 - 2 P_{2}{}\bigr)^2 \bigl(2725 {P_{1}}^4 - 14708 {P_{1}}^2 P_{2}{} + 19860 {P_{2}}^2\bigr)  \nonumber\\
	&&-  \tfrac{256}{3} {a_{2}}^8 \bigl(27145 {P_{1}}^9 - 255384 {P_{1}}^7 P_{2}{} + 893304 {P_{1}}^5 {P_{2}}^2 - 1378784 {P_{1}}^3 {P_{2}}^3 + 793104 P_{1}{} {P_{2}}^4\bigr)\Bigr).\nonumber
\end{eqnarray}
We aim to derive the flow equations pertinent to this overarching Lagrangian in~\eqref{Lpp}. Such equations will conform to the archetype of standard irrelevant  $T \bar{T}$  deformations within a two-dimensional framework, applicable to theories characterized by integer powers of the Lorentz Invariant variables $ P_1 $ and $ P_2 $. To accomplish this, we commence by determining the energy-momentum tensor from the general Lagrangian, denoted as Eq.~\eqref{Lpp}, derived in the following manner:
\begin{eqnarray}\label{Tmn}
	T_{\mu \nu }&=& {\cal L} \,\,g_{\mu \nu }+ {\cal F}(P_1,P_2, \lambda)\,\, \partial_{\mu }\Phi^{f} \partial_{\nu }\Phi_{f} + {\cal G} (P_1,P_2, \lambda)\,\, \partial_{\alpha }\Phi_{g} \partial^{\alpha }\Phi_{f} \partial_{\mu }\Phi^{f} \partial_{\nu }\Phi^{g}.
\end{eqnarray}
Appendix~\eqref{Appenn} contains the details of the two functions ${\cal F}(P_1,P_2, \lambda)$ and ${\cal G} (P_1,P_2, \lambda)$, specifically in Eqs.~\eqref{FF} and~\eqref{GG}.
In the context of the various theories governed by PDE~\eqref{lagrangeNGZ2D}, the flow equation is contingent upon two specific variables, denoted as $T_{\mu \nu } T^{\mu \nu }$ and $T^{\mu}{}_{ \mu } T^{\nu}{}_{ \nu }$. To elucidate, the flow equation is generally represented by the formula $\frac{\partial{{\cal L}(\lambda)_{NGZ}}}{\partial{\lambda}}=f(T_{\mu \nu } T^{\mu \nu }, T^{\mu}{}_{ \mu } T^{\nu}{}_{ \nu })$. For the derivation of $f(T_{\mu \nu } T^{\mu \nu }, T^{\mu}{}_{ \mu } T^{\nu}{}_{ \nu })$, it is imperative to construct two foundational structures:
$T_{\mu \nu } T^{\mu \nu }$ 
and
$T^{\mu}{}_{ \mu } T^{\nu}{}_{ \nu }$.
The configuration of $T_{\mu \nu } T^{\mu \nu }$ can be deduced from~\eqref{Tmn} via the equation:
\begin{eqnarray}\label{TTabNGZ}
	T_{\mu \nu } T^{\mu \nu } &=&- \tfrac{1}{2} {P_{1}}^2 + P_{2}{} + \lambda a_{2}{} \bigl(-4 {P_{1}}^3 + 8 P_{1}{} P_{2}{}\bigr) + \lambda^2 {a_{2}}^2 \bigl(-14 {P_{1}}^4 + 8 {P_{1}}^2 P_{2}{} + 40 {P_{2}}^2\bigr)\\
	&& + \lambda^3 \Bigl(2 a_{7}{} P_{1}{} \bigl({P_{1}}^2 - 2 P_{2}{}\bigr)^2 + 64 {a_{2}}^3 \bigl({P_{1}}^5 - 8 {P_{1}}^3 P_{2}{} + 12 P_{1}{} {P_{2}}^2\bigr)\Bigr) \nonumber\\
	&&+ \lambda^4 \Bigl(a_{2}{} a_{7}{} \bigl({P_{1}}^2 - 2 P_{2}{}\bigr)^2 \bigl(29 {P_{1}}^2 + 22 P_{2}{}\bigr) \nonumber\\
	&&+ 32 {a_{2}}^4 \bigl(39 {P_{1}}^6 - 174 {P_{1}}^4 P_{2}{} + 148 {P_{1}}^2 {P_{2}}^2 + 88 {P_{2}}^3\bigr)\Bigr)\nonumber\\
	&& + \lambda^5 \Bigl(-8 {a_{2}}^2 a_{7}{} P_{1}{} \bigl(17 {P_{1}}^2 - 154 P_{2}{}\bigr) \bigl({P_{1}}^2 - 2 P_{2}{}\bigr)^2 + 2 a_{13}{} P_{1}{} \bigl({P_{1}}^2 - 2 P_{2}{}\bigr)^3\nonumber\\
	&& + 128 {a_{2}}^5 \bigl(13 {P_{1}}^7 + 146 {P_{1}}^5 P_{2}{} - 836 {P_{1}}^3 {P_{2}}^2 + 984 P_{1}{} {P_{2}}^3\bigr)\Bigr)\nonumber\\
	&& + \lambda^6 \Bigl(\tfrac{2}{3} a_{2}{} a_{13}{} \bigl({P_{1}}^2 - 2 P_{2}{}\bigr)^3 \bigl(67 {P_{1}}^2 + 34 P_{2}{}\bigr) -  \tfrac{1}{8} {a_{7}}^2 \bigl({P_{1}}^2 - 2 P_{2}{}\bigr)^3 \bigl(87 {P_{1}}^2 + 50 P_{2}{}\bigr)\nonumber\\
	&& -  \tfrac{40}{3} {a_{2}}^3 a_{7}{} \bigl({P_{1}}^2 - 2 P_{2}{}\bigr)^2 \bigl(563 {P_{1}}^4 - 1496 {P_{1}}^2 P_{2}{} - 604 {P_{2}}^2\bigr)\nonumber\\
	&& -  \tfrac{128}{3} {a_{2}}^6 \bigl(3173 {P_{1}}^8 - 23200 {P_{1}}^6 P_{2}{} + 51960 {P_{1}}^4 {P_{2}}^2 - 28288 {P_{1}}^2 {P_{2}}^3 - 16432 {P_{2}}^4\bigr)\Bigr)\nonumber\\
	&& + \lambda^7 \Bigl(32 a_{2}{} {a_{7}}^2 P_{1}{} \bigl(3 {P_{1}}^2 - 34 P_{2}{}\bigr) \bigl({P_{1}}^2 - 2 P_{2}{}\bigr)^3 \nonumber\\
	&&-  \tfrac{256}{3} {a_{2}}^2 a_{13}{} P_{1}{} \bigl(2 {P_{1}}^2 - 25 P_{2}{}\bigr) \bigl({P_{1}}^2 - 2 P_{2}{}\bigr)^3 \nonumber\\
	&&+ 2 a_{21}{} P_{1}{} \bigl({P_{1}}^2 - 2 P_{2}{}\bigr)^4 -  \tfrac{512}{3} {a_{2}}^4 a_{7}{} P_{1}{} \bigl({P_{1}}^2 - 2 P_{2}{}\bigr)^2 \bigl(11 {P_{1}}^4 + 1594 {P_{1}}^2 P_{2}{} - 4072 {P_{2}}^2\bigr) \nonumber\\
	&& + \tfrac{2048}{3} {a_{2}}^7 \bigl(59 {P_{1}}^9- 7936 {P_{1}}^7 P_{2}{} + 51768 {P_{1}}^5 {P_{2}}^2 - 114496 {P_{1}}^3 {P_{2}}^3 + 84464 P_{1}{} {P_{2}}^4\bigr)\Bigr).\nonumber
\end{eqnarray}
To compute the aforementioned configuration, the relationships
\begin{eqnarray}\label{iden}
	&&\partial_{\alpha}\Phi^{j} \partial^{\alpha}\Phi^{i} \partial_{\beta}\Phi^{k} \partial^{\beta}\Phi_{i} \partial_{\gamma}\Phi_{k} \partial^{\gamma}\Phi_{j}=- \frac{1}{2}\, P_1^3 + \frac{3}{2}\, P_1\, P_2\,,\\
	&&\partial_{\alpha }\Phi^{g} \partial^{\alpha }\Phi^{f} \partial_{\beta }\Phi^{h} \partial^{\beta }\Phi_{f} \partial_{\gamma }\Phi^{i} \partial^{\gamma }\Phi_{g} \partial_{\delta }\Phi_{i} \partial^{\delta }\Phi_{h}=- \tfrac{1}{2}{ P_{1}}^4 +{ P_{1}}^2 P_{2}{} + \tfrac{1}{2}{ P_{2}}^2,\nonumber
\end{eqnarray}
have been employed.
The structure of $T^{\mu}{}_{ \mu } T^{\nu}{}_{ \nu }$ can be directly derived from~\eqref{Tmn}, as delineated below:
\begin{eqnarray}\label{TaaTbbNGZ}
	T^{\mu}{}_{ \mu } T^{\nu}{}_{ \nu } &=&4 \lambda^2 {a_{2}}^2 \bigl({P_{1}}^2 - 2 P_{2}{}\bigr)^2 + 64 \lambda^3 {a_{2}}^3 P_{1}{} \bigl({P_{1}}^2 - 2 P_{2}{}\bigr)^2\\
	&&  + \lambda^4 \Bigl(-6 a_{2}{} a_{7}{} \bigl({P_{1}}^2 - 2 P_{2}{}\bigr)^3 + 64 {a_{2}}^4 \bigl({P_{1}}^2 - 2 P_{2}{}\bigr)^2 \bigl(7 {P_{1}}^2 + 6 P_{2}{}\bigr)\Bigr) \nonumber\\
	&&+ \lambda^5 \Bigl(-176 {a_{2}}^2 a_{7}{} P_{1}{} \bigl({P_{1}}^2 - 2 P_{2}{}\bigr)^3 + 10240 {a_{2}}^5 P_{1}{} \bigl({P_{1}}^2 - 2 P_{2}{}\bigr)^2 P_{2}{}\Bigr)\nonumber\\
	&&+ \lambda^6 \Bigl(\tfrac{9}{4} {a_{7}}^2 \bigl({P_{1}}^2 - 2 P_{2}{}\bigr)^4 -  \tfrac{20}{3} a_{2}{} a_{13}{} \bigl({P_{1}}^2 - 2 P_{2}{}\bigr)^4 \nonumber\\
	&&-  \tfrac{16}{3} {a_{2}}^3 a_{7}{} \bigl({P_{1}}^2 - 2 P_{2}{}\bigr)^3 \bigl(319 {P_{1}}^2 + 454 P_{2}{}\bigr)\nonumber\\
	&& -  \tfrac{256}{3} {a_{2}}^6 \bigl({P_{1}}^2 - 2 P_{2}{}\bigr)^2 \bigl(197 {P_{1}}^4 - 620 {P_{1}}^2 P_{2}{} - 1228 {P_{2}}^2\bigr)\Bigr) \nonumber\\
	&&+ \lambda^7 \Bigl(\tfrac{128}{3} {a_{2}}^4 a_{7}{} P_{1}{} \bigl(251 {P_{1}}^2 - 2182 P_{2}{}\bigr) \bigl({P_{1}}^2 - 2 P_{2}{}\bigr)^3 + 144 a_{2}{} {a_{7}}^2 P_{1}{} \bigl({P_{1}}^2 - 2 P_{2}{}\bigr)^4 \nonumber\\
	&&-  \tfrac{736}{3} {a_{2}}^2 a_{13}{} P_{1}{} \bigl({P_{1}}^2 - 2 P_{2}{}\bigr)^4 \nonumber\\
	&&+ \tfrac{2048}{3} {a_{2}}^7 P_{1}{} \bigl({P_{1}}^2 - 2 P_{2}{}\bigr)^2 \bigl(415 {P_{1}}^4 - 3004 {P_{1}}^2 P_{2}{} + 5692 {P_{2}}^2\bigr)\Bigr)\nonumber 
.
\end{eqnarray}
In our previous work, detailed in \cite{Babaei-Aghbolagh:2024jho}, we established that within a four-dimensional framework, the general form of irrelevant  $T \bar{T}$-like deformations for actions characterized by integral exponents of the Lorentzian variables $ t$ and $z$ manifests as a sequence comprising dual constructs: $  T_{\mu \nu } T^{\mu \nu } $ and $ T_{\mu }{}^{\mu }{} T_{\nu }{}^{\nu }$. This pattern is also discernible within a two-dimensional context for the overarching Lagrangian labeled as~\eqref{Lpp}. Consequently, we can deduce the expression for this 
general perturbation, articulated as the ensuing general flow equation, accurate to the order of $  \lambda^7 $:
\begin{align}\label{GTTbarSeri2D}
	\boxed{ \frac{\partial {\cal L}(\lambda)}{\partial \lambda}=\sum_{n=0}^{\infty} c_n \frac{(T_{\mu }{}^{\mu }{} T_{\nu }{}^{\nu }{})^n}{( T_{\mu \nu } T^{\mu \nu })^{n-1}}}
\end{align}
with coefficients $c_n$ as:
\begin{eqnarray}\label{GTTbarExpandXY}
	&&c_0=-2 a_{2},\,\,\,\,\,c_1=- \frac{16 {a_{2}}^3 + 3 a_{7}{}}{16 {a_{2}}^2},\,\,\,\,\,  \\
	&&c_2= - \frac{50176 {a_{2}}^6 + 1664 {a_{2}}^3 a_{7}{} - 27 {a_{7}}^2 - 20 a_{2}{} a_{13}{}}{768 {a_{2}}^5}, \nonumber\\
	&& c_3=- \frac{121044992 {a_{2}}^9 + 4264960 {a_{2}}^6 a_{7}{} - 40512 {a_{2}}^3 {a_{7}}^2  }{49152 {a_{2}}^8}\nonumber \\
	&& - \frac{ 567 {a_{7}}^3 - 38656 {a_{2}}^4 a_{13}{} + 720 a_{2}{} a_{7}{} a_{13}{} + 168 {a_{2}}^2 a_{21}{}}{49152 {a_{2}}^8}.\nonumber
\end{eqnarray}
Flow equation~\eqref{GTTbarSeri2D}  provides a general perturbative formulation of flow dynamics in $D=2$, capturing a broad class of theoretical models.
with the appropriate exponents of $P_1$ and $P_2$, as they pertain to PDE~\eqref{lagrangeNGZ2D}. This particular flow equation aligns with the overarching principles of the general Lagrangian, denoted as equation~\eqref{Lpp}.

%%%%%%%%%%%%%%%%%%%%%%%%%%%%%%%%%%%%%%%%%%%%%%%%%%%%%%%%%%%%%%%%%%%%%%%%%%%%%%%%%%%%%%%%%%%%%%%%%%%%%%%%%%%%%%%%%%%%%%%%%%%%%%%%%%%%%%%%%%%%%%%%%%%%%%%%%%%%%%%%%%%%%%%%%%%%%%%%%%%%%%%%%%%%%%%%%%%%%%%%%%%%
\section{General root $T\bar{T}$-like flows in two dimensions}\label{Sec3}
%%%%%%%%%%%%%%%%%%%%%%%%%%%%%%%%%%%%%%%%%%%%%%%%%%%%%%%%%%%%%%%%%%%%%%%%%%%%%%%%%%%%%%%%%%%%%%%%%%%%%%%%%%%%%%%%%%%%%%%%%%%%%%%%%%%%%%%%%%%%%%%%%%%%%%%%%%%%%%%%%%%%%%%%%%%%%%%%%%%%%%%%%%%%%%%%%%%%%
The duality-invariant PDE~\eqref{lagrangeNGZ2D} can be simplified by expressing $\mathcal{L}$  as a function of $P_1$ and
\begin{eqnarray}\label{DMt}
	\mathcal{P} := \sqrt{- {P_{1}}^2 + 2 P_{2}{}} \,.
\end{eqnarray}
The duality-invariant PDE  for $\mathcal{L}(P_1, \mathcal{P} )$ is
\begin{eqnarray}\label{DEP1P}
	4 \bigl(\frac{\partial \mathcal{L}}{\partial P_1}\bigr)^2-4 \bigl(\frac{\partial \mathcal{L}}{\partial \mathcal{P}}\bigr)^2-1 =0\,.
\end{eqnarray}
Upon implementing the variable substitution~\eqref{DMt}, the GSMM Lagrangian in~\eqref{GSMMdw} is reformulated thus:
\begin{eqnarray}\label{GMSP}
	\mathcal{L}_{GSMM} =\frac{2}{ \lambda}\Big( 1 -  \sqrt{1 + \frac{1}{2} \lambda \big( \cosh(\gamma)\, P_1 +   \sinh(\gamma)\, \mathcal{P} \big)+\frac{1}{16} \lambda^2 ( P_1^2- \mathcal{P}^2 ) }\Big).
\end{eqnarray}
It can be explicitly verified that Lagrangian~\eqref{GMSP} satisfies PDE~\eqref{DEP1P}.
A similar approach to the previous section can solve self-duality PDE~\eqref{DEP1P} with a series of integer powers of the two independent variables, $P_1$ and $\mathcal{P}$. In all the theories we consider, there are only two independent invariants $P_1$ and $\mathcal{P}$. All higher-order invariants are combinations of these two. For this purpose, we consider the following generic Lagrangian:
\begin{eqnarray}\label{LPP1}
	\mathcal{L}(\lambda, \gamma) &=&d_1 \mathcal{P} +d_2 P_1 + \lambda (d_3 \mathcal{P}^2 + d_4 \mathcal{P} P_1 + d_5 P_1^2) + \lambda^2 (d_6 \mathcal{P}^3 +d_7 \mathcal{P}^2 P_1 + d_8 \mathcal{P} P_1^2 + d_9 P_1^3)  \nonumber\\
	&&+ \lambda^3 (d_{10} \mathcal{P}^4 +d_{11} \mathcal{P}^3 P_1 + d_{12} \mathcal{P}^2 P_1^2 + d_{13} \mathcal{P} P_1^3 + d_{14} P_1^4)\nonumber\\
	&& + \lambda^4 (d_{15} \mathcal{P}^5 + d_{16} \mathcal{P}^4 P_1 + d_{17} \mathcal{P}^3 P_1^2 +d_{18} \mathcal{P}^2 P_1^3 + d_{19} \mathcal{P} P_1^4 + d_{20} P_1^5).
\end{eqnarray}
By employing a perturbation approach similar to the previous section, we can solve PDE~\eqref{DEP1P} using the general solution~\eqref{LPP1}, step by step. These conditions are essential for solving Eq.~\eqref{DEP1P} up to the $\lambda$ order in the form of $\Big( d_{2}{} \to - \tfrac{1}{2} \sqrt{1+4 {d_{1}}^2}\Big)\,\,\,\,\, ,\,\,\,\,\,\, \Big(d_{3 }\to -\frac{ d_{4}{}}{4 {d_{1}}} \sqrt{1+4 {d_{1}}^2}\,\,,\,\,\,d_{5} \to - \frac{d_{1} d_{4}}{\sqrt{1+4 {d_{1}}^2}}\Big)$. In this scenario, only one unknown coefficient $d_n$ remains at each order of $\lambda$. The perturbed solutions of PDE~\eqref{DEP1P}, assuming $d_1=- \tfrac{1}{2} \sinh(\gamma) $, up to order of $\lambda^4$ are as follows:
\begin{eqnarray}\label{Lmar}
	\mathcal{L}(\lambda, \gamma)&=&-\tfrac{1}{2} \bigl( P_1 \cosh(\gamma) +  \mathcal{P} \sinh(\gamma)\bigr) + \lambda \,d_4 \frac{ \bigl(P_1 + \mathcal{P} \coth(\gamma)\bigr)^2}{2  \coth(\gamma)} \\
	&&+ \lambda^2 \Bigl(\tfrac{1}{3} d_4^2 \bigl(\mathcal{P}  \csch(\gamma) - 2 P_1 \sech (\gamma)\bigr) \bigl(\mathcal{P}  \csch(\gamma) + P_1  \sech(\gamma)\bigr)^2\nonumber\\
	&& + \frac{d_7 \bigl(P_1 + \mathcal{P}  \coth(\gamma)\bigr)^3}{3 \bigl( \coth(\gamma)\bigr)^2}\Bigr)\nonumber\\
	&& + \lambda^3 \biggl(\frac{d_4 d_7 \bigl(\mathcal{P}  \csch(\gamma) - 3 P_1  \sech(\gamma)\bigr) \bigl(\mathcal{P}  \csch(\gamma) + P_1  \sech(\gamma)\bigr)^3}{2  \csch(\gamma)} \nonumber\\
	&&+ \frac{d_{11} \bigl(P_1 + \mathcal{P}  \coth(\gamma)\bigr)^4}{4 \bigl( \coth(\gamma)\bigr)^3}\nonumber\\
	&&+ \frac{d_4^3 \bigl(\mathcal{P}  \csch(\gamma) + P_1  \sech(\gamma)\bigr)^2 \Bigl(\bigl(P_1 -  \mathcal{P}  \coth(\gamma)\bigr)^2 \bigl( \csch(\gamma)\bigr)^2 - 4 P_1^2 \bigl( \sech(\gamma)\bigr)^2\Bigr)}{2  \coth(\gamma)}\biggr) \nonumber\\
	&& + \lambda^4 \Biggr(\frac{2 d_{11} d_4 \bigl(\mathcal{P}  \csch(\gamma) - 4 P_1  \sech(\gamma)\bigr) \bigl(\mathcal{P}  \csch(\gamma) + P_1  \sech(\gamma)\bigr)^4}{5 \bigl( \csch(\gamma)\bigr)^2} 
	\nonumber\\
	&& + \frac{d_7^2 \bigl(\mathcal{P}  \csch(\gamma) - 4 P_1  \sech(\gamma)\bigr) \bigl(\mathcal{P}  \csch(\gamma) + P_1  \sech(\gamma)\bigr)^4}{5 \bigl( \csch(\gamma)\bigr)^2}  + \frac{d_{16} \bigl(P_1 + \mathcal{P}  \coth(\gamma)\bigr)^5}{5 \bigl( \coth(\gamma)\bigr)^4} 
	\nonumber\\
	&& + \tfrac{1}{20} d_4^2 d_7 \bigl( \csch(\gamma)\bigr)^5 \bigl( \sech(\gamma)\bigr)^6 \bigl(\mathcal{P} \cosh(\gamma) + P_1 \sinh(\gamma)\bigr)^3  
	\nonumber\\
	&& \times \Bigl(-68 P_1^2 + 9 \mathcal{P}^2 + 4 (20 P_1^2 + 3 \mathcal{P}^2) \cosh(2 \gamma) + 3 (-4 P_1^2 + \mathcal{P}^2) \cosh(4 \gamma)\nonumber\\
	&& - 72 P_1 \mathcal{P} \bigl(\cosh(\gamma)\bigr)^3 \sinh(\gamma)\Bigr)  
	\nonumber\\
	&& + d_4^4 \Biggl(\mathcal{P}^5 \bigl( \csch(\gamma)\bigr)^7 + \bigl( \csch(\gamma)\bigr)^5 \Bigl(\tfrac{4}{5} \mathcal{P}^5 - 2 P_1^2 \mathcal{P}^3 \bigl( \sech(\gamma)\bigr)^4\Bigr)\nonumber\\
	&& + \bigl( \csch(\gamma)\bigr)^3 \Bigl(-6 P_1^2 \mathcal{P}^3 \bigl( \sech(\gamma)\bigr)^4 + P_1^4 \mathcal{P} \bigl( \sech(\gamma)\bigr)^6\Bigr)\nonumber\\
	&& +  \csch(\gamma) \Bigl(-8 P_1^2 \mathcal{P}^3 \bigl( \sech(\gamma)\bigr)^4 + 9 P_1^4 \mathcal{P} \bigl( \sech(\gamma)\bigr)^6\Bigr)\nonumber\\
	&& + \tfrac{4}{5} P_1^3 \bigl( \sech(\gamma)\bigr)^5 \biggl(-4 (P_1^2 + 5 \mathcal{P}^2) + 5 P_1 \Bigl(2 P_1 \bigl( \sech(\gamma)\bigr)^2 - 3 \mathcal{P} \tanh(\gamma)\Bigr)\biggr)\Biggr)\Biggr).\nonumber
\end{eqnarray}
The most general Lagrangian, denoted $\mathcal{L}(\lambda, \gamma)$ in~\eqref{Lmar}, can be constructed in the presence of two coupling constants $\lambda$ and $\gamma$, up to the order $\lambda^4$. In the above Lagrangian, the coefficients of $d_i(\gamma)$  depend exclusively on this dependent function of  $\gamma$ . The  coefficients dependent on $\gamma$ in this Lagrangian can be systematically determined by application of the root flow equation originating from the structure of the Lagrangian coefficients. For this end, we first obtain the momentum energy tensor of this general integrable theory.
The energy-momentum tensor corresponding to the general Lagrangian~\eqref{Lmar} can be determined as follows:
\begin{eqnarray}\label{Tmng}
	T_{\mu \nu }={\cal L}(\lambda,  \gamma) g_{\mu \nu }+ {\cal H}(P_1,\mathcal{P}, \lambda, \gamma) \partial_{\mu }\Phi^{f} \partial_{\nu }\Phi_{f} + {\cal U} (P_1,\mathcal{P}, \lambda, \gamma) \partial_{\alpha }\Phi_{g} \partial^{\alpha }\Phi_{f} \partial_{\mu }\Phi^{f} \partial_{\nu }\Phi^{g}.
\end{eqnarray}
The functions ${\cal H}(P_1,\mathcal{P}, \lambda,\, \gamma)$ and ${\cal U}(P_1,\mathcal{P}, \lambda,\, \gamma)$ are defined in Appendix~\eqref{Appenn}  , in Eqs.~\eqref{HH} and~\eqref{UU}. 
To derive a root flow equation, we construct two configurations, $T_{\mu \nu} T^{\mu \nu}$ and $T^{\mu}{}_{\mu} T^{\nu}{}_{\nu}$, from the energy-momentum tensor presented in Eq.~\eqref{Tmng}. The configurations for these two structures are provided in Appendix~\eqref{Appenn}, in Eqs.~\eqref{TTabkga} and~\eqref{TaaTbbfa}.
We aim to study the root flow equation within the framework of the general Lagrangian~\eqref{Lmar}. We consider a general form of the root flow equation derived from the irrelevant flow equation~\eqref{GTTbarSeri2D}. Consequently, the general Lagrangian~\eqref{Lmar} must satisfy the following flow equation:
\begin{align}\label{GRoot2D}
	\boxed{\frac{\partial {\cal L}(\lambda,\gamma )}{\partial \gamma}= 
		\sqrt{ \sum_{n=0}^{\infty} e_n \frac{(T_{\mu }{}^{\mu }{} T_{\nu }{}^{\nu }{})^n}{( T_{\mu \nu } T^{\mu \nu })^{n-1}}}\,.}
\end{align}
The general Lagrangian~\eqref{Lmar} simplifies to the (SMM) theory in~\eqref{SMMdw}  in the limit of $\lambda=0$. Notably, there is no $\gamma$-dependent coefficient. The solution to root flow equation~\eqref{GRoot2D} is derived with the constant $e_0=\tfrac{1}{2}$.
In the next order of $\lambda$, we encounter the unknown constant $d_4$, which determines the $\gamma$-dependence of $d_4$ from the solution of root flow equation~\eqref{GRoot2D}. To achieve this, we substitute~\eqref{TTabkga},~\eqref{TaaTbbfa}, and~\eqref{Lmar} into root flow equation~\eqref{GRoot2D} and simplify in the  order of $\lambda$, resulting in the following differential equation:
\begin{align}\label{eqd4}
	2 \cosh(2\gamma){d}_4   - \sinh(2\gamma) {d}'_4  = 0.
\end{align}
Solving differential equation~\eqref{eqd4} yields the $\gamma$-dependence of the coefficient $d_4$ as follows:
\begin{align}\label{d4}
	{d}_4 =  n_{1}{\cosh(\gamma)} \sinh(\gamma) ,
\end{align}
where $n_1$ is a constant independent of $\gamma$. By applying this method to the order of $\lambda^2$, solving root flow equation~\eqref{GRoot2D} results in a differential equation that depends on the coefficient $d_{7}$ and ${d}'_7$. The differential equation is as follows:
\begin{eqnarray}\label{eqd7}
	&&n_1^2 \bigl( 16 \cosh(2 \gamma) + 3 (1 + 4 e_1) \cosh(4 \gamma)-51 - 12e_1\bigr)\nonumber\\
	&& + d_7 \bigl( 64  \sech(\gamma)-96 \cosh(\gamma) \bigr) + 32 {d}'_7 \sinh(\gamma)  = 0.
\end{eqnarray}
By solving differential equation~\eqref{eqd7}, we obtain the $\gamma$-dependence of the coefficient $d_7$ as follows:
\begin{align}\label{dn7}
	{d}_7 = \frac{n_2 \sqrt{\cosh(\gamma)} \bigl(\sinh(2 \gamma)\bigr)^{3/2}}{\sqrt{\sinh(\gamma)}} -  \tfrac{1}{8} n_1^2 \cosh(\gamma) \bigl(-4 + 12 \cosh(2 \gamma) + 3 (1 + 4 e_1) \gamma \sinh(2 \gamma)\bigr).
\end{align}
Using this method for the order of $\lambda^3$, the $d_{11}$ coefficient is determined as follows:
\begin{eqnarray}\label{d11}
	{d}_{11} &= & \tfrac{1}{2} n_1^3 \cosh(\gamma) \bigl(3 (1 + 4 e_1) \gamma \cosh(3 \gamma)\\
	&& - 8 \sinh(\gamma)\bigr) + \bigl(n_3 + n_1^3 (\gamma + 4 e_1 \gamma)^2\bigr) \bigl(\cosh(\gamma)\bigr)^3 \sinh(\gamma) \nonumber\\
	&& -  \frac{2 n_1 n_2 \sqrt{\cosh(\gamma)} \bigl(\sinh(2 \gamma)\bigr)^{3/2} \bigl(-3 + 6 \cosh(2 \gamma) + 2 (1 + 4 e_1) \gamma \sinh(2 \gamma)\bigr)}{3 \bigl(\sinh(\gamma)\bigr)^{3/2}}.\nonumber
\end{eqnarray}
At the $\lambda^4$ order, the differential equation arising from the root deformation~\eqref{GRoot2D} is quite lengthy. We solve this equation for the specific case where $e_1 = -\frac{1}{4}$, and determine the coefficient $d_{16}$ as follows:
\begin{eqnarray}\label{d16}
	{d}_{16} &= & (-4 n_2^2 -  n_1 n_3) \bigl(\cosh(\gamma)\bigr)^3 \bigl(5 \cosh(2 \gamma)-3 \bigr) \\
	&&+ \tfrac{1}{8} n_1^4 \cosh(\gamma) \bigl(-29 + 100 \cosh(2 \gamma) + 65 \cosh(4 \gamma)\bigr) \nonumber\\
	&& - 48 n_1^2 n_2 \bigl(\cosh(\gamma)\bigr)^{3/2} \bigl(\sinh(\gamma)\bigr)^{1/2} \bigl(\sinh(2 \gamma)\bigr)^{1/2}  + \frac{n_4 \bigl(\sinh(2 \gamma)\bigr)^{5/2}}{\bigl(\tanh(\gamma)\bigr)^{3/2}} \nonumber\\
	&&-  \frac{10 n_1^4 e_2 \gamma \bigl(\cosh(\gamma)\bigr)^{5/2} \bigl( \coth(\tfrac{1}{2} \gamma)\bigr)^{1/2} \bigl(\sinh(\gamma)\bigr)^{5/2} \bigl(\tanh(\tfrac{1}{2} \gamma)\bigr)^{1/2}}{\bigl(\tanh(\gamma)\bigr)^{3/2}}.\nonumber
\end{eqnarray}
In this section, we employed the perturbation approach to examine the general form of two-dimensional scalar theories up to order $\lambda^4$. We expressed this as the Lagrangian in~\eqref{LPP1} and identified constraints on the coefficients $d_n$ to satisfy the $SO(N)\times SO(N)$  duality invariant condition in the differential form~\eqref{DEP1P}. This process led us to derive the Lagrangian in~\eqref{Lmar}. Additionally, we determined the $\gamma$ dependence of the unfixed coefficients $d_n$ by imposing the constraint that the theory adheres to a general root flow equation of the form~\eqref{GRoot2D}. By substituting these coefficients~\eqref{d4},~\eqref{dn7},~\eqref{d11}, and~\eqref{d16} into the Lagrangian in~\eqref{Lmar}, we obtained the most general form of the Lagrangian with two coupling constants up to order $\lambda^4$, ensuring that it is both $SO(N)\times SO(N)$  duality invariant and consistent with the root flow equation.

%%%%%%%%%%%%%%%%%%%%%%%%%%%%%%%%%%%%%%%%%%%%%%%%%%%%%
\section{Generalized irrelevant flow equation and commutability}\label{0409}
%%%%%%%%%%%%%%%%%%%%%%%%%%%%%%%%%%%%%%%%%%%%%%%%%%%%
%\subsubsection*{\underline{ \it{A specific choice:  }}}
Given the coefficients $e_n$ as $e_0=\frac{1}{2}$,  $e_1 = -\frac{1}{4}$, and $e_2=e_3=...=e_n=0$, the root flow equation in~\eqref{GRoot2D} will take the standard form of~\eqref{Margi} in two dimensions. Now, with this choice for $e_n$ and the substitution of the coefficient $d_n$ in the general Lagrangian~\eqref{Lmar}, we obtain a general Lagrangian that satisfies the root flow equation $\frac{\partial \mathcal{L}(\lambda, \gamma)}{\partial \gamma}=\mathcal{R}_\gamma$. This Lagrangian is as follows:
\begin{eqnarray}\label{LLL}
	\mathcal{L}(\lambda, \gamma) &&= -\tfrac{1}{2} \bigl( P_1 \cosh(\gamma) +  \mathcal{P}  \sinh(\gamma)\bigr)+ \tfrac{1}{2} n_1 \lambda \bigl(\mathcal{P}  \cosh(\gamma) + P_1 \sinh(\gamma)\bigr)^2 \\
	&&  + \lambda^2 \Bigl(- n_1^2 \bigl(\mathcal{P}  \cosh(\gamma) + P_1 \sinh(\gamma)\bigr)^2 \bigl(P_1 \cosh(\gamma) + \mathcal{P}  \sinh(\gamma)\bigr) \nonumber\\
	&& + \frac{2 n_2 \bigl(\mathcal{P}  \cosh(\gamma) + P_1 \sinh(\gamma)\bigr)^3 \bigl(\sinh(2 \gamma)\bigr)^{1/2}}{3 \bigl(\cosh(\gamma)\bigr)^{1/2} \bigl(\sinh(\gamma)\bigr)^{1/2}}\Bigr) \nonumber\\
	&& + \lambda^3 \Bigl(-2 n_1^3 (- P_1 + \mathcal{P} ) (P_1 + \mathcal{P} ) \bigl(\mathcal{P}  \cosh(\gamma) + P_1 \sinh(\gamma)\bigr)^2\nonumber\\
	&& + \tfrac{1}{4} n_3 \bigl(\mathcal{P}  \cosh(\gamma) + P_1 \sinh(\gamma)\bigr)^4\nonumber\\
	&&  -  \frac{4 n_1 n_2 \bigl(\mathcal{P}  \cosh(\gamma) + P_1 \sinh(\gamma)\bigr)^3 \bigl(P_1 \cosh(\gamma) + \mathcal{P}  \sinh(\gamma)\bigr) \bigl(\sinh(2 \gamma)\bigr)^{1/2}}{\bigl(\cosh(\gamma)\bigr)^{1/2} \bigl(\sinh(\gamma)\bigr)^{1/2}}\Bigr)\nonumber\\
	&&  + \lambda^4 \Bigl(-8 n_2^2 \bigl(\mathcal{P}  \cosh(\gamma) + P_1 \sinh(\gamma)\bigr)^4 \bigl(P_1 \cosh(\gamma) + \mathcal{P}  \sinh(\gamma)\bigr)\nonumber\\
	&&  - 2 n_1 n_3 \bigl(\mathcal{P}  \cosh(\gamma) + P_1 \sinh(\gamma)\bigr)^4 \bigl(P_1 \cosh(\gamma) + \mathcal{P}  \sinh(\gamma)\bigr)\nonumber\\
	&&  -  \frac{16 n_1^2 n_2 (- P_1 + \mathcal{P} ) (P_1 + \mathcal{P} ) \bigl(\mathcal{P}  \cosh(\gamma) + P_1 \sinh(\gamma)\bigr)^3 \bigl(\sinh(2 \gamma)\bigr)^{1/2}}{\bigl(\cosh(\gamma)\bigr)^{1/2} \bigl(\sinh(\gamma)\bigr)^{1/2}}\nonumber\\
	&&  + \tfrac{1}{4} n_1^4 \bigl(\mathcal{P}  \cosh(\gamma) + P_1 \sinh(\gamma)\bigr)^2 \nonumber\\
	&& \times \bigl(29 P_1 (- P_1 + \mathcal{P} ) (P_1 + \mathcal{P} ) \cosh(\gamma) + 13 P_1 (P_1^2 + 3 \mathcal{P} ^2) \cosh(3 \gamma)\nonumber\\
	&& + 29 \mathcal{P}  (- P_1 + \mathcal{P} ) (P_1 + \mathcal{P} ) \sinh(\gamma) + 13 \mathcal{P}  (3 P_1^2 + \mathcal{P} ^2) \sinh(3 \gamma)\bigr) \nonumber\\
	&& + \frac{128 n_4 \bigl(\cosh(\gamma)\bigr)^2 \bigl(\sinh(\gamma)\bigr)^7 \bigl(\mathcal{P}  \cosh(\gamma) + P_1 \sinh(\gamma)\bigr)^5}{5 \bigl(\sinh(2 \gamma)\bigr)^{9/2} \bigl(\tanh(\gamma)\bigr)^{5/2}}\Bigr) \nonumber.
\end{eqnarray}
We can explicitly verify that the Lagrangian~\eqref{LLL} possesses a root flow equation. With respect to root deformation, the general Lagrangian in~\eqref{LLL} simplifies to the two-dimensional ModMax Lagrangian when $\lambda=0$. This general Lagrangian includes an unfixed coefficient $n_i$ on each order of $\lambda$. Specifically, on the order $\lambda^4$, it contains four unfixed coefficients: $n_1$, $n_2$, $n_3$, and $n_4$. We can derive theories that exhibit a root flow equation by determining these constant coefficients. For example, setting the constants $n_i$ to $n_1=\frac{1}{8}$, $n_2=0$, $n_3=\frac{5}{256}$, and $n_4=0$, we can expand the general two-dimensional Generalized Scalar ModMax (GSMM) theory in Eq.~\eqref{GSMMdw} up to the order $\lambda^4$.

The general integrable theory derived in Eq.~\eqref{LLL} contains an undetermined constant at each order of the $\lambda$-expansion, where these coefficients are theory-dependent. As established in the literature, this framework enables the dimensional reduction of four-dimensional electrodynamic theories to two dimensions, with Born-Infeld theory serving as a prime example. Crucially, this reduction preserves causality when applied to four-dimensional theories that respect the causality principle, producing novel two-dimensional integrable systems (see \cite{russo2024causal, russo2024dual} for representative cases). The resulting two-dimensional theories maintain closed-form expressions, and their $\lambda$- expansions are systematically captured by our general Lagrangian in Eq.~\eqref{LLL}. Specific theories emerge from this unified framework through the appropriate choices of the expansion coefficients. This construction represents a significant advancement as it provides: 1: a perturbative classification of integrable theories with two field variables ($P_1$, $P_2$), 2: consistent incorporation of two coupling constants ($\gamma$, $\lambda$), and a solid foundation for future investigations in this field.

This section will discuss the importance of identifying the general irrelevant flow equation for Lagrangian~\eqref{LLL} that commutes with the root deformation in~\eqref{Margi}.
By taking the derivative of Lagrangian~\eqref{LLL} with respect to $\lambda$ and comparing it with structures~\eqref{TTabkga} and~\eqref{TaaTbbfa}, we can derive the  flow equation of the Lagrangian~\eqref{LLL} using a perturbation approach up to the $\lambda^4$ order as follows:
\begin{eqnarray}\label{GFE}
	\frac{\partial \mathcal{L}(\lambda, \gamma)}{\partial \lambda} =\sum_{m=0}^{\infty} C_m Y^{\tfrac{m}{2}}  X^{1-\tfrac{m}{2}}  \,,
\end{eqnarray}
where $X=T_{\mu\nu}T^{\mu\nu}$ and $Y={T_{\mu}}^{\mu}{T_{\nu}}^{\nu}$ and coefficients $C_m$ as:
\begin{eqnarray}\label{GTXY}
	&&C_0= n_{1},\,\,\,\,\,C_1=\frac{64}{3} n_2,\,\,\,\,\,C_2= -\frac{17}{16}-\frac{32768}{9} n_2^2+48 \, n_3,\\
	&& C_3=\frac{16}{135} \big(-1125 \, n_2 + 10485760 \, n_2^3 - 207360 \, n_2 \, n_3 + 13824 \, n_4\big).\nonumber
\end{eqnarray}
The irrelevant flow equation~\eqref{GFE} is a general flow equation applicable to all theories with arbitrary $ n_i$. The coefficients $ C_m $ of this equation explicitly depend on the coefficients of $n_i$. We explicitly derive the irrelevant $T \bar{T}$-like deformation of the Lagrangian~\eqref{LLL} in~\eqref{GFE}, which represents the general irrelevant  $T \bar{T}$-like deformation. This deformation commutes with the root deformation in~\eqref{Margi} as follows:
\begin{align}\label{CO}
	\boxed{\frac{\partial \mathcal{L}(\lambda, \gamma)}{\partial \lambda} =\sum_{m=0}^{\infty} C_m Y^{\tfrac{m}{2}}  X^{1-\tfrac{m}{2}} \,,\,\,\,\,\,\,\,\,\,\,\,\,\frac{\partial  \mathcal{L}(\lambda, \gamma)}{\partial \gamma}=\frac{1}{\sqrt{2}}\, \sqrt{X-\frac{1}{2} Y} }\,.
\end{align}
In this paper, we shown that operators ${\cal O}_{\lambda}=\sum_{m=0}^{\infty} C_m Y^{\tfrac{m}{2}}  X^{1-\tfrac{m}{2}} $ and  ${\cal R}_{\gamma}$ exhibit commutativity.  Consequently, it can be explicitly verified that the {\it{double-flow equation:}} $\partial_\lambda \partial_\gamma  \mathcal{L}(\lambda, \gamma)$, holds for the Lagrangian~\eqref{LLL} up to the order of $\lambda^4$ in the identity of $\partial_\lambda \partial_\gamma  \mathcal{L}(\lambda, \gamma)-\partial_\gamma \partial_\lambda\mathcal{L}(\lambda, \gamma)=0$. This property is illustrated in Fig.~\eqref{fig2}.
\begin{center}
	\begin{tikzcd}[row sep=huge,column sep=huge]
		\mathcal{L}(0,0) \arrow[r, blue, "{\cal O}_{\lambda}" blue] \arrow[d,red,"{\cal R}_{\gamma}" red]
		&|[blue]| \mathcal{L}(\lambda, 0) \arrow[d,red, "{\cal R}_{\gamma}" red] \\
		|[red]|\mathcal{L}(0, \gamma) \arrow[r, blue, "{\cal O}_{\lambda}" blue]
		&|[red!50!blue]|  \mathcal{L}(\lambda, \gamma)
	\end{tikzcd}
	\captionof{figure}{Deformations of the general  multi-scalar theories under ${\cal O}_{\lambda}$ and ${\cal R}_{\gamma}$.}\label{fig2}
\end{center}

The Lagrangian of~\eqref{LLL} can be categorized into two classes, each characterized by irrelevant $T \bar{T}$-like deformations. We can consider $n_i$ as $i=even$ and $i=odd$. The first class includes $n_i$ where $n_{even}=0$   and $n_{odd} \ne 0$ . These theories exhibit irrelevant flow equations with integer powers of the energy-momentum tensor.
The second class consists of theories with $n_{even} \ne 0$   and $n_{odd}\ne 0$,  featuring flow equations with fractional powers of the energy-momentum tensor as~\eqref{GFE}. Interestingly, the first class is a special subgroup of the second class. In other words, when we study theories involving $n_{even}=0$   and $n_{odd} \ne 0$, it automatically transforms into $C_{odd}=0$, leaving only the integer powers of the two structures, $X$ and $Y$.

%\paragraph{Class I ($n_{even}=0$   and $n_{odd} \ne 0$):}
%\paragraph{Class II ($n_{even} \ne 0$   and $n_{odd} \ne 0$):}
%%%%%%%%%%%%%%%%%%%%%%%%%%%%%%%%%%%%%%%%%%%%%%%%%%%%%
\section{Conclusions and perspectives}\label{04}
%%%%%%%%%%%%%%%%%%%%%%%%%%%%%%%%%%%%%%%%%%%%%%%%%%%%
In this work, we investigate the relationship between $T\bar{T}$ deformations (both root and irrelevant) and scalar field theories, with a focus on the $SO(N) \times SO(N)$ symmetry and integrability inherent in these systems. We demonstrate that the necessary condition for duality invariance under the $SO(N) \times SO(N)$ symmetry group manifests as a PDE in two-dimensional scalar theories. This PDE corresponds precisely to the integrability condition, a crucial element for the consistency and solvability of the theory.

The irrelevant $T\bar{T}$ deformation, initially introduced in the context of two-dimensional quantum field theories in \cite{Smirnov:2016lqw, Cavaglia:2016oda}, has garnered significant attention for its ability to preserve the integrability of the theory. This leads to a rich structure of deformed models. We derive a general perturbation solution to this PDE, encompassing both the root $T\bar{T}$ flow equation and a general irrelevant $T\bar{T}$-like flow equation. These equations describe the theory's evolution under deformation and are essential for understanding the modified dynamics.

The resulting theory in~\eqref{LLL} exhibits three fundamental properties: first, it contains theory-dependent coefficients at each order of the $\lambda$-expansion; second,
it proposes a perturbative framework for classifying integrable systems 
with two Lorentz-invariant variables ($P_1$, $P_2$) for arbitrary couplings ($\lambda$, $\gamma$); and third, it unifies known and novel integrable theories under a single consistent formalism. This work thereby outlines a constructive method and suggests a perturbative classification scheme for this class of theories.

Furthermore, we identify a general form for these flow equations that commute with each other. This non-trivial commutativity implies that the order in which the deformations are applied does not affect the outcome, which is essential for maintaining duality invariance. The commutative nature of flow equations underpins the robustness of the theoretical framework and opens new avenues for further research in this area of theoretical physics. Potential applications of these results extend to theories of chiral $p-$forms  \cite{Bandos:2020hgy, Mkrtchyan:2019opf, Bansal:2021bis,Avetisyan:2021heg, Avetisyan:2022zza}, supersymmetry \cite{Bandos:2021rqy, Ferko:2023wyi}, and gravitational theories involving the Ricci flow equations \cite{Brizio2024c, Morone2024c}. In \cite{ Morone2024c}, it is briefly discussed that the general marginal deformation ~\eqref{GRoot2D} at the   Lagrangian level corresponds to a "pure change of metric."

Our findings also show that generalized $T\bar{T}$ deformations in two-dimensional scalar field theories align with corresponding scalar theories derived from irrelevant flow behavior obtained through reduction of the dimensionality of duality-invariant nonlinear electrodynamics in four dimensions \cite{Conti:2018jho, Babaei2024c}. The proposed deformation framework provides a comprehensive understanding of the behavior of both irrelevant and marginal deformations, leading to a generalized theory in lower dimensions.

The generalized theory unifies different deformation paths governed by two coupled constants, $\lambda$ and $\gamma$, each controlling various aspects of the deformation structure. Importantly, our results reveal the rich underlying nature of these models. Future research could explore higher-order solutions and their implications for integrability, quantum consistency, and potential connections to string theory via compactification mechanisms. In addition, it is intriguing to identify an operator derived from the energy-momentum tensor that characterizes  {\it{double-flow equation}}.
We also aim to derive closed-form expressions for the general Lagrangians discussed in this paper in future work. One approach to obtaining closed-form Lagrangians involves the Courant-Hilbert method. The PDE~\eqref{lagrangeNGZ2D} reduces, through appropriate changes of variables, to the Courant-Hilbert equation. The most general solution to this equation is given by the Courant-Hilbert function $\ell(\tau)$, which depends on a parameterized real variable $\tau$.  An alternative perspective on this equivalence can be obtained using the auxiliary field formalism introduced in \cite{Ferko2024cw}.

\section*{Acknowledgments}
The authors are grateful to Dmitri Sorokin, Roberto Tateo and Karapet Mkrtchyan for valuable discussions.  S.H. acknowledges financial support from Ningbo University, the Max Planck Partner Group, and the Natural Science Foundation of China Grants No.~12475053, No.~12075101, No.~12235016, No.~12347209. H.O. is supported by the National Natural Science Foundation of China, Grant No. 12205115, and by the Science and Technology Development Plan Project of Jilin Province of China, Grant No. 20240101326JC.	

%%%%%%%%%%%%%%%%%%%%%%%%%%%%%%%%%%%
%%%%%%%%%%%%%%%%%%%%%%%%%%%%%%%%%%%%%%%%%%%%%%%%%%%%%%%%%%%%%%%%%%%%%%%%%%%%%%%%%%%%%%%%%%%%%%%%%%%%%%%%%%%%%%%%%%%%%%%%%%%%%%%%%%%%%%%%%%%%%%%%%%%%%%%%%%%%%%%%%%%%%%%%%%%%%%%%%%%%%%%%%%%%%%%%%%%%%%%%%%%%%%%%%%%%%%%%%%%%%%%%%%%%%%%%%%%%%%%%%%%%%%%%%%%%%%%%%%%%%%%%%%%%%%%%%%%%%%%%%%%%%%%%%%%%%%%%%%%%%%%%%%%%%
\appendix

\section{Details of the Lagrangian, and the Energy-Momentum Tensor}\label{Appenn}
This appendix provides a detailed account of some of the perturbation calculations from the paper, which are extensive.  Given that general Lagrangian~\eqref{lagrangeGeneral} is valid in the PDE~\eqref{lagrangeNGZ2D}, there exists a relationship between the coefficients $a_n$ at each order of $\lambda$, resulting in the following correlation between the coefficients $a_n$ as:

\begin{eqnarray}
	\label{an}
	&&\Big(a_{1} \to - \tfrac{1}{2} \Big);\,\,\,\,\,\,\,\,\Big(a_{3}\to -2 a_{2} \Big);\,\,\,\,\,\,\,\,\Big(a_{4}\to 4 {a_{2}}^2 ,a_{5}\to -8 {a_{2}}^2 \Big);\,\,\,\,\,\,\,\,\\
	&&\Big(a_{6}{} \to \tfrac{1}{4} \bigl(32 {a_{2}}^3 -  a_{7}{}\bigr)\,\,,\,\,a_{8}{}\to -32 {a_{2}}^3 -  a_{7}{}\Big);\nonumber\\
	&&\Big(a_{10}{}\to 16 \bigl(20 {a_{2}}^4 + a_{2}{} a_{7}{}\bigr)\,\,,\,\,a_{11}\to -16 \bigl(28 {a_{2}}^4 + a_{2}{} a_{7}{}\bigr)\,\,,\,\,a_{9}\to -4 \bigl(12 {a_{2}}^4 + a_{2}{} a_{7}{}\bigr)\Big);\nonumber\\
	&&\Big(a_{12}{} \to  \tfrac{1}{6} \bigl(-1024 {a_{2}}^5 - 80 {a_{2}}^2 a_{7}{} -  a_{13}{}\bigr)\,\,,\,\,a_{14}\to -2 \bigl(-1536 {a_{2}}^5 - 80 {a_{2}}^2 a_{7}{} + a_{13}{}\bigr)\,\,,\,\,\nonumber\\
	&&a_{15}\to \tfrac{4}{3} \bigl(-3584 {a_{2}}^5 - 160 {a_{2}}^2 a_{7}{} + \
	a_{13}{}\bigr)\Big);\,\,\,\,\,\,\,\,\Big(a_{16}\to 7808 {a_{2}}^6 + 352 {a_{2}}^3 a_{7}{} + {a_{7}}^2 - 4 a_{2} a_{13}{}\,\,,\,\,\nonumber\\
	&& a_{17}\to 2 \bigl(-29056 {a_{2}}^6 - 1376 {a_{2}}^3 a_{7}{} - 3 {a_{7}}^2 \
	+ 12 a_{2}{} a_{13}{}\bigr)\,\,,\,\, \nonumber\\
	&&a_{18}\to -4 \bigl(-35712 {a_{2}}^6 - 1696 {a_{2}}^3 a_{7}{} - 3 {a_{7}}^2 + 12 a_{2}{} a_{13}{}\bigr)\nonumber\\
	&&a_{19}\to 8 \bigl(-14464 {a_{2}}^6 - 672 {a_{2}}^3 a_{7}{} -  {a_{7}}^2 +	4 a_{2}{} a_{13}{}\bigr) \Big);\nonumber\\
	&&\Big(a_{20}\to  \tfrac{1}{8} \bigl(229376 {a_{2}}^7 + 10752 {a_{2}}^4 a_{7}{} + 56 a_{2}{} {a_{7}}^2 - 112 {a_{2}}^2 a_{13}{} -  a_{21}{}\bigr),\nonumber\\
	&&a_{22} \to -843776 {a_{2}}^7 - 41216 {a_{2}}^4 a_{7}{} - 168 a_{2}{} \
	{a_{7}}^2 + 336 {a_{2}}^2 a_{13}{} - 3 a_{21}{},\nonumber\\
	&&a_{23}\to -4 \bigl(-622592 {a_{2}}^7 - 30464 {a_{2}}^4 a_{7}{} - 112 a_{2}{} {a_{7}}^2 + 224 {a_{2}}^2 a_{13}{} -  a_{21}{}\bigr),\nonumber\\
	&&,a_{24} \to 2 \bigl(-1032192 {a_{2}}^7 - 50176 {a_{2}}^4 a_{7}{} - 168 a_{2}{} {a_{7}}^2 + 336 {a_{2}}^2 a_{13}{} -  a_{21}{}\bigr)  \Big) \nonumber\\
	&&\Big(a_{25}\to \tfrac{2}{3} \bigl(-3474560 {a_{2}}^8 - 174400 {a_{2}}^5 a_{7}{} \
	- 780 {a_{2}}^2 {a_{7}}^2 + 1168 {a_{2}}^3 a_{13}{} + 3 a_{7}{} \
	a_{13}{} - 6 a_{2}{} a_{21}{}\bigr)\nonumber\\
	&& a_{26}\to -16 \bigl(-1362048 {a_{2}}^8 - 68288 {a_{2}}^5 a_{7}{} - 316 \
	{a_{2}}^2 {a_{7}}^2 + 464 {a_{2}}^3 a_{13}{} + a_{7}{} a_{13}{} \
	- 2 a_{2}{} a_{21}{}\bigr)\nonumber\\
	&& a_{27}\to 16 \bigl(-4764288 {a_{2}}^8 - 238912 {a_{2}}^5 a_{7}{} - 1116 \
	{a_{2}}^2 {a_{7}}^2 + 1616 {a_{2}}^3 a_{13}{} + 3 a_{7}{} \
	a_{13}{} - 6 a_{2}{} a_{21}{}\bigr) \nonumber\\
	&& a_{28} \to - \tfrac{64}{3} \bigl(-5515136 {a_{2}}^8 - 276544 {a_{2}}^5 \
	a_{7}{} - 1284 {a_{2}}^2 {a_{7}}^2 + 1840 {a_{2}}^3 a_{13}{} + \
	3 a_{7}{} a_{13}{} - 6 a_{2}{} a_{21}{}\bigr)  \nonumber\\
	&& a_{29} \to 32 \bigl(-2114944 {a_{2}}^8 - 105920 {a_{2}}^5 a_{7}{} - 484 {a_{2}}^2 {a_{7}}^2 + 688 {a_{2}}^3 a_{13}{} + a_{7}{} a_{13}{} - 2 a_{2}{} a_{21}{}\bigr)\Big).\nonumber
\end{eqnarray}
The details of the energy-momentum tensor are provided in Eq.~\eqref{Tmn}, involving two functions, $ {\cal F}(P_1,P_2, \lambda)$ and ${\cal G}(P_1,P_2, \lambda)$, as follows:
\begin{eqnarray}\label{FF}
	&&{\cal F}(P_1,P_2, \lambda) =1 - 4 \lambda a_{2}{} P_{1}{} + 8 \lambda^2 {a_{2}}^2 \bigl(-3 {P_{1}}^2 + 2 P_{2}{}\bigr) + \lambda^3 \Bigl(2 \bigl(-32 {a_{2}}^3 + a_{7}{}\bigr) {P_{1}}^3 - 4 a_{7}{} P_{1}{} P_{2}{}\Bigr)\nonumber \\
	&& + 8 \lambda^4 a_{2}{} \Bigl(5 \bigl(12 {a_{2}}^3 + a_{7}{}\bigr) {P_{1}}^4 - 12 \bigl(20 {a_{2}}^3 + a_{7}{}\bigr) {P_{1}}^2 P_{2}{} + 4 \bigl(28 {a_{2}}^3 + a_{7}{}\bigr) {P_{2}}^2\Bigr) \nonumber\\
	&&+ \lambda^5 \biggl(2 a_{13}{} P_{1}{} \bigl({P_{1}}^2 - 2 P_{2}{}\bigr)^2 + 32 {a_{2}}^2 P_{1}{} \Bigl(64 {a_{2}}^3 \bigl({P_{1}}^4 - 6 {P_{2}}^2\bigr) + 5 a_{7}{} \bigl({P_{1}}^4 - 4 {P_{2}}^2\bigr)\Bigr)\biggr) \nonumber\\
	&& + \lambda^6 \biggl(8 a_{2}{} a_{13}{} \bigl({P_{1}}^2 - 2 P_{2}{}\bigr)^2 \bigl(7 {P_{1}}^2 - 2 P_{2}{}\bigr) - 2 \Bigl({a_{7}}^2 \bigl({P_{1}}^2 - 2 P_{2}{}\bigr)^2 \bigl(7 {P_{1}}^2 - 2 P_{2}{}\bigr) \nonumber\\
	&& + 128 {a_{2}}^6 \bigl(427 {P_{1}}^6 - 2270 {P_{1}}^4 P_{2}{} + 3348 {P_{1}}^2 {P_{2}}^2 - 904 {P_{2}}^3\bigr) \nonumber\\
	&& + 32 {a_{2}}^3 a_{7}{} \bigl(77 {P_{1}}^6 - 430 {P_{1}}^4 P_{2}{} + 636 {P_{1}}^2 {P_{2}}^2 - 168 {P_{2}}^3\bigr)\Bigr)\biggr) \nonumber\\
	&& + 2 \lambda^7 P_{1}{} \Bigl(a_{21}{} \bigl({P_{1}}^2 - 2 P_{2}{}\bigr)^3 - 56 a_{2}{} {a_{7}}^2 \bigl({P_{1}}^2 - 2 P_{2}{}\bigr)^2 \bigl({P_{1}}^2 + 4 P_{2}{}\bigr) \nonumber\\
	&& + 112 {a_{2}}^2 a_{13}{} \bigl({P_{1}}^2 - 2 P_{2}{}\bigr)^2 \bigl({P_{1}}^2 + 4 P_{2}{}\bigr)  - 3584 {a_{2}}^4 a_{7}{} \bigl(3 {P_{1}}^6 - 46 {P_{1}}^2 {P_{2}}^2 + 68 {P_{2}}^3\bigr) \nonumber\\
	&&- 32768 {a_{2}}^7 \bigl(7 {P_{1}}^6 - 103 {P_{1}}^2 {P_{2}}^2 + 152 {P_{2}}^3\bigr)\Bigr)+{\cal O}( \lambda^8). 
\end{eqnarray}
and
\begin{eqnarray}\label{GG}
	&&{\cal G}(P_1,P_2, \lambda) 
	=8 \lambda a_{2}{} + 32 \lambda^2 {a_{2}}^2 P_{1}{}  + \lambda^3 \Bigl(-4 a_{7}{} {P_{1}}^2 + 8 \bigl(32 {a_{2}}^3 + a_{7}{}\bigr) P_{2}{}\Bigr) \\
	&& + 64 \lambda^4 a_{2}{} P_{1}{} \Bigl(\bigl(-20 {a_{2}}^3 -  a_{7}{}\bigr) {P_{1}}^2 + 2 \bigl(28 {a_{2}}^3 + a_{7}{}\bigr) P_{2}{}\Bigr) \nonumber\\
	&&+ \lambda^5 \biggl(-4 a_{13}{} \bigl({P_{1}}^2 - 2 P_{2}{}\bigr)^2 - 256 {a_{2}}^2 \Bigl(32 {a_{2}}^3 \bigl(3 {P_{1}}^2 - 7 P_{2}{}\bigr) + 5 a_{7}{} \bigl({P_{1}}^2 - 2 P_{2}{}\bigr)\Bigr) P_{2}{}\biggr) \nonumber\\
	&& + 8 \lambda^6 P_{1}{} \Bigl(3 {a_{7}}^2 \bigl({P_{1}}^2 - 2 P_{2}{}\bigr)^2 - 12 a_{2}{} a_{13}{} \bigl({P_{1}}^2 - 2 P_{2}{}\bigr)^2  \nonumber\\
	&&+ 32 {a_{2}}^3 a_{7}{} \bigl(43 {P_{1}}^4 - 212 {P_{1}}^2 P_{2}{} + 252 {P_{2}}^2\bigr)\nonumber\\
	&& + 128 {a_{2}}^6 \bigl(227 {P_{1}}^4 - 1116 {P_{1}}^2 P_{2}{} + 1356 {P_{2}}^2\bigr)\Bigr) \nonumber\\
	&& + \lambda^7 \biggl(-4 a_{21}{} \bigl({P_{1}}^2 - 2 P_{2}{}\bigr)^3 + 64 a_{2}{} P_{2}{} \Bigl(21 {a_{7}}^2 \bigl({P_{1}}^2 - 2 P_{2}{}\bigr)^2 - 42 a_{2}{} a_{13}{} \bigl({P_{1}}^2 - 2 P_{2}{}\bigr)^2  \nonumber\\
	&&+ 224 {a_{2}}^3 a_{7}{} \bigl(23 {P_{1}}^4 - 102 {P_{1}}^2 P_{2}{} + 112 {P_{2}}^2\bigr)\nonumber\\
	&& + 1024 {a_{2}}^6 \bigl(103 {P_{1}}^4 - 456 {P_{1}}^2 P_{2}{} + 504 {P_{2}}^2\bigr)\Bigr)\biggr)+{\cal O}( \lambda^8). \nonumber
\end{eqnarray}
The Eq.~\eqref{Tmng} provides the details of the energy-momentum tensor, expressed through the functions $ {\cal H}$ and $ {\cal U} $ as follows:
\begin{eqnarray}\label{HH}
	&&{\cal H}=\cosh(\gamma) -  \frac{P_1 \sinh(\gamma)}{ \mathcal{P}} + d_4 \lambda \bigl(\frac{2 P_1^2}{ \mathcal{P}} - 2  \mathcal{P} + 2 P_1  \csch(\gamma)  \sech(\gamma)\bigr) \\
	&&+ \lambda^2 \biggl(\frac{2 d_4^2 P_1 \bigl( \mathcal{P}  \csch(\gamma) + P_1  \sech(\gamma)\bigr) \Bigl( \mathcal{P} \bigl( \csch(\gamma)\bigr)^2 -  P_1  \csch(\gamma)  \sech(\gamma) + 2  \mathcal{P} \bigl( \sech(\gamma)\bigr)^2\Bigr)}{ \mathcal{P}} \nonumber\\
	&&+ \frac{2 d_7 \bigl(-  \mathcal{P} + P_1  \coth(\gamma)\bigr) \bigl( \mathcal{P} + P_1 \tanh(\gamma)\bigr)^2}{ \mathcal{P}}\biggr)\nonumber\\
	&&+ \lambda^3 \biggl(- \frac{4 d_4 d_7 P_1 \bigl(P_1 +  \mathcal{P}  \coth(\gamma)\bigr)^2  \csch(\gamma) \bigl( \sech(\gamma)\bigr)^4 \bigl( \mathcal{P} - 2  \mathcal{P} \cosh(2 \gamma) + P_1 \sinh(2 \gamma)\bigr)}{ \mathcal{P}}\nonumber\\
	&& + \frac{2 d_{11} \bigl(-  \mathcal{P} + P_1  \coth(\gamma)\bigr) \bigl( \mathcal{P} + P_1 \tanh(\gamma)\bigr)^3}{ \mathcal{P}} \nonumber\\
	&&+ \frac{4 d_4^3 P_1 }{ \mathcal{P}}\Bigl( \mathcal{P}^3  \coth(\gamma) \bigl( \csch(\gamma)\bigr)^4 +  \mathcal{P} (- P_1 +  \mathcal{P}) (P_1 +  \mathcal{P}) \bigl( \csch(\gamma)\bigr)^3  \sech(\gamma)\nonumber\\
	&& +  \mathcal{P} (-3 P_1^2 + 2  \mathcal{P}^2)  \csch(\gamma) \bigl( \sech(\gamma)\bigr)^3 - 2 P_1 \bigl( \sech(\gamma)\bigr)^4 \bigl(P_1^2 - 3  \mathcal{P}^2 - 2 P_1  \mathcal{P} \tanh(\gamma)\bigr)\Bigr)\biggr)\nonumber\\
	&& + \lambda^4 \Biggl(\frac{2 d_{11} d_4 P_1 \bigl(P_1 +  \mathcal{P}  \coth(\gamma)\bigr)^3 \bigl( \sech(\gamma)\bigr)^5 \Bigl(5  \mathcal{P} \cosh(2 \gamma) - 3 \bigl( \mathcal{P} + P_1 \sinh(2 \gamma)\bigr)\Bigr)}{ \mathcal{P}} \nonumber\\
	&& + \frac{d_7^2 P_1 \bigl(P_1 +  \mathcal{P}  \coth(\gamma)\bigr)^3 \bigl( \sech(\gamma)\bigr)^5 \Bigl(5  \mathcal{P} \cosh(2 \gamma) - 3 \bigl( \mathcal{P} + P_1 \sinh(2 \gamma)\bigr)\Bigr)}{ \mathcal{P}} \nonumber\\
	&&  + \frac{2 d_{16} \bigl(-  \mathcal{P} + P_1  \coth(\gamma)\bigr) \bigl( \mathcal{P} + P_1 \tanh(\gamma)\bigr)^4}{ \mathcal{P}}\nonumber\\
	&& + \frac{d_4^2 d_7 P_1 \bigl( \csch(\gamma)\bigr)^5 \bigl( \sech(\gamma)\bigr)^6 \bigl( \mathcal{P} \cosh(\gamma) + P_1 \sinh(\gamma)\bigr)^2 }{4  \mathcal{P}}\nonumber\\
	&& \times \bigl((-30 P_1^2 + 38  \mathcal{P}^2) \cosh(\gamma) + (39 P_1^2 - 5  \mathcal{P}^2) \cosh(3 \gamma) + 3 (-3 P_1^2 + 5  \mathcal{P}^2) \cosh(5 \gamma)\nonumber\\
	&& + 204 P_1  \mathcal{P} \sinh(\gamma) - 110 P_1  \mathcal{P} \sinh(3 \gamma) + 6 P_1  \mathcal{P} \sinh(5 \gamma)\bigr)
	\nonumber\\
	&& + \frac{2 d_4^4 P_1 }{ \mathcal{P}}\Biggl( \csch(\gamma) \biggl(P_1^4 \Bigl(6 + \bigl( \csch(\gamma)\bigr)^2\Bigr) - 2 P_1^2  \mathcal{P}^2 \Bigl(15 + 5 \bigl( \csch(\gamma)\bigr)^2 + 3 \bigl( \csch(\gamma)\bigr)^4\Bigr)\nonumber\\
	&& +  \mathcal{P}^4 \Bigl(4 + 4 \bigl( \csch(\gamma)\bigr)^2 + 8 \bigl( \csch(\gamma)\bigr)^4 + 5 \bigl( \csch(\gamma)\bigr)^6\Bigr)\biggr) + 16 P_1  \mathcal{P} (- P_1^2 + 3  \mathcal{P}^2) \bigl( \sech(\gamma)\bigr)^5 \nonumber\\
	&&- 40 P_1^3  \mathcal{P} \bigl( \sech(\gamma)\bigr)^7 - 2 (3 P_1^4 - 15 P_1^2  \mathcal{P}^2 + 2  \mathcal{P}^4)  \sech(\gamma) \tanh(\gamma) \nonumber\\
	&&+ (-7 P_1^4 + 40 P_1^2  \mathcal{P}^2 - 8  \mathcal{P}^4) \bigl( \sech(\gamma)\bigr)^3 \tanh(\gamma) - 20 P_1^2 (P_1^2 - 4  \mathcal{P}^2) \bigl( \sech(\gamma)\bigr)^5 \tanh(\gamma)\Biggr)\Biggr) \nonumber
\end{eqnarray}
and
\begin{eqnarray}\label{UU}
	&&{\cal U} =\frac{2 \sinh(\gamma)}{\mathcal{P}} -\lambda \frac{4 d_4  \bigl( \coth(\gamma) \mathcal{P}+ P_1\bigr)}{\mathcal{P}}\\
	&& + \lambda^2 \biggl(d_4^2 \Bigl(-4 \bigl( \csch(\gamma)\bigr)^3 \mathcal{P}+ \frac{4  \csch(\gamma) P_1^2 \bigl( \sech(\gamma)\bigr)^2}{\mathcal{P}}\Bigr) -  \frac{4 d_7 \bigl( \coth(\gamma) \mathcal{P}+ P_1\bigr)^2}{ \coth(\gamma) \mathcal{P}}\biggr)\nonumber\\
	&&+ \lambda^3 \Biggl(- \frac{8 d_4 d_7 \bigl( \csch(\gamma) \mathcal{P}- 2 P_1  \sech(\gamma)\bigr) \bigl( \csch(\gamma) \mathcal{P}+ P_1  \sech(\gamma)\bigr)^2}{\mathcal{P}}-  \frac{4 d_{11} \bigl( \coth(\gamma) \mathcal{P}+ P_1\bigr)^3}{\bigl( \coth(\gamma)\bigr)^2 P}\nonumber\\
	&& + d_4^3 \biggl(-8  \coth(\gamma) \bigl( \csch(\gamma)\bigr)^4 \mathcal{P}^2 \nonumber\\
	&&+ \frac{8 P_1^2  \sech(\gamma) \Bigl(\bigl( \csch(\gamma)\bigr)^3 \mathcal{P}+ 2  \csch(\gamma) \mathcal{P}\bigl( \sech(\gamma)\bigr)^2 + 2 P_1 \bigl( \sech(\gamma)\bigr)^3\Bigr)}{\mathcal{P}}\biggr) \Biggr) \nonumber\\
	&&+ \lambda^4 \Biggl(- \frac{8 d_{11} d_4 \bigl( \csch(\gamma) \mathcal{P}- 3 P_1  \sech(\gamma)\bigr) \bigl( \csch(\gamma) \mathcal{P}+ P_1  \sech(\gamma)\bigr)^3}{ \csch(\gamma) \mathcal{P}}\nonumber\\
	&& -  \frac{4 d_7^2 \bigl( \csch(\gamma) \mathcal{P}- 3 P_1  \sech(\gamma)\bigr) \bigl( \csch(\gamma) \mathcal{P}+ P_1  \sech(\gamma)\bigr)^3}{ \csch(\gamma) \mathcal{P}} -  \frac{4 d_{16} \bigl( \coth(\gamma) \mathcal{P}+ P_1\bigr)^4}{\bigl( \coth(\gamma)\bigr)^3 P} \nonumber\\
	&&-  \frac{3 d_4^2 d_7 \bigl( \csch(\gamma)\bigr)^5 \bigl( \sech(\gamma)\bigr)^5 \bigl(\cosh(\gamma) \mathcal{P}+ P_1 \sinh(\gamma)\bigr)^2 }{\mathcal{P}}\nonumber\\
	&& \times\Bigl(3 \mathcal{P}^2 - 13 P_1^2 + \cosh(4 \gamma) (\mathcal{P}^2 - 3 P_1^2) + 4 \cosh(2 \gamma) (\mathcal{P}^2 + 4 P_1^2) - 16 \bigl(\cosh(\gamma)\bigr)^3 \mathcal{P}P_1 \sinh(\gamma)\Bigr)\nonumber\\
	&& + \frac{4 d_4^4 }{\mathcal{P}} \biggl( 6  \csch(\gamma) (\mathcal{P}-  P_1) P_1^2 (\mathcal{P}+ P_1)-5 \bigl( \csch(\gamma)\bigr)^7 \mathcal{P}^4  \nonumber\\
	&&-  \bigl( \csch(\gamma)\bigr)^3 P_1^2 ( P_1^2-6 \mathcal{P}^2 )  + \bigl( \csch(\gamma)\bigr)^5 ( 6 \mathcal{P}^2 P_1^2-4 \mathcal{P}^4)\nonumber\\
	&& + P_1^2  \sech(\gamma) \Bigl(6 ( P_1^2- \mathcal{P}^2 ) \tanh(\gamma) + ( 7 P_1^2-12 \mathcal{P}^2 ) \bigl( \sech(\gamma)\bigr)^2 \tanh(\gamma) \nonumber\\
	&&+ 4 P_1 \bigl( \sech(\gamma)\bigr)^4 \bigl(8 \mathcal{P}+ 5 P_1 \tanh(\gamma)\bigr)\Bigr)\biggr)\Biggr),\nonumber
\end{eqnarray}
By substituting Eqs.~\eqref{HH} and~\eqref{UU} into the energy-momentum tensor~\eqref{Tmng}, we can easily derive the two structures, $	T_{\mu \nu } T^{\mu \nu }$ and $T^{\mu}{}_{ \mu } T^{\nu}{}_{ \nu }$. The details of these structures are as follows:
\begin{eqnarray}\label{TTabkga}
	&&T_{\mu \nu } T^{\mu \nu } =\tfrac{1}{2} \bigl(\mathcal{P} \cosh(\gamma) + P_1 \sinh(\gamma)\bigr)^2\nonumber\\
	&&  - 2\lambda  d_4   \csch(\gamma)  \sech(\gamma) \bigl(\mathcal{P} \cosh(\gamma) + P_1 \sinh(\gamma)\bigr)^2 \bigl(P_1 \cosh(\gamma) + \mathcal{P} \sinh(\gamma)\bigr)\nonumber\\
	&&   + \lambda^2 \Bigl(-2 d_7  \csch(\gamma) \bigl( \sech(\gamma)\bigr)^2 \bigl(\mathcal{P} \cosh(\gamma) + P_1 \sinh(\gamma)\bigr)^3 \bigl(P_1 \cosh(\gamma) + \mathcal{P} \sinh(\gamma)\bigr)\nonumber\\
	&&  + \tfrac{1}{16} d_4^2 \bigl( \csch(\gamma)\bigr)^3 \bigl( \sech(\gamma)\bigr)^3 \bigl(\mathcal{P} \cosh(\gamma) + P_1 \sinh(\gamma)\bigr)^2 \bigl(-58 P_1 \mathcal{P} + 16 P_1 \mathcal{P} \cosh(2 \gamma) \nonumber\\
	&& + 10 P_1 \mathcal{P} \cosh(4 \gamma) + 2 (11 P_1^2 - 3 \mathcal{P}^2) \sinh(2 \gamma) + 5 (P_1^2 + \mathcal{P}^2) \sinh(4 \gamma)\bigr)\Bigr) \nonumber\\
	&& + \lambda^3 \Bigl(-2 d_{11}  \csch(\gamma) \bigl( \sech(\gamma)\bigr)^3 \bigl(\mathcal{P} \cosh(\gamma) + P_1 \sinh(\gamma)\bigr)^4 \bigl(P_1 \cosh(\gamma) + \mathcal{P} \sinh(\gamma)\bigr) \nonumber\\
	&& -  \tfrac{2}{3} d_4 d_7  \sech(\gamma) \bigl(\mathcal{P}  \csch(\gamma) + P_1  \sech(\gamma)\bigr)^3 \bigl(14 P_1 \mathcal{P} - 6 P_1 \mathcal{P} \cosh(2 \gamma) - 2 P_1 \mathcal{P} \cosh(4 \gamma) \nonumber\\
	&& + (-7 P_1^2 + \mathcal{P}^2) \sinh(2 \gamma) -  (P_1^2 + \mathcal{P}^2) \sinh(4 \gamma)\bigr)\nonumber\\
	&&  + \tfrac{1}{12} d_4^3 \bigl( \csch(\gamma)\bigr)^5 \bigl( \sech(\gamma)\bigr)^5 \bigl(\mathcal{P} \cosh(\gamma) + P_1 \sinh(\gamma)\bigr)^2 \bigl(14 P_1 (P_1^2 - 3 \mathcal{P}^2) \cosh(\gamma) \nonumber\\
	&& - 3 P_1 (5 P_1^2 + 3 \mathcal{P}^2) \cosh(3 \gamma) + P_1 (P_1^2 + 3 \mathcal{P}^2) \cosh(5 \gamma) + 2 \mathcal{P} (81 P_1^2 + \mathcal{P}^2) \sinh(\gamma) \nonumber\\
	&& + 3 \mathcal{P} (-9 P_1^2 + \mathcal{P}^2) \sinh(3 \gamma) + \mathcal{P} (3 P_1^2 + \mathcal{P}^2) \sinh(5 \gamma)\bigr)\Bigr)\nonumber\\
	&&  + \lambda^4 \Bigl(-2 d_{16}  \csch(\gamma) \bigl( \sech(\gamma)\bigr)^4 \bigl(\mathcal{P} \cosh(\gamma) + P_1 \sinh(\gamma)\bigr)^5 \bigl(P_1 \cosh(\gamma) + \mathcal{P} \sinh(\gamma)\bigr)\nonumber\\
	&&  + \tfrac{1}{144} d_4^2 d_7 \bigl( \csch(\gamma)\bigr)^5 \bigl( \sech(\gamma)\bigr)^6 \bigl(\mathcal{P} \cosh(\gamma) + P_1 \sinh(\gamma)\bigr)^3 \bigl(2 P_1 (607 P_1^2 - 765 \mathcal{P}^2) \cosh(\gamma)\nonumber\\
	&&  - 15 P_1 (83 P_1^2 + 21 \mathcal{P}^2) \cosh(3 \gamma) + P_1 (31 P_1^2 + 117 \mathcal{P}^2) \cosh(5 \gamma) + 2 \mathcal{P} (4863 P_1^2 + 43 \mathcal{P}^2) \sinh(\gamma)\nonumber\\
	&&  + 3 \mathcal{P} (-563 P_1^2 + 43 \mathcal{P}^2) \sinh(3 \gamma) + \mathcal{P} (105 P_1^2 + 43 \mathcal{P}^2) \sinh(5 \gamma)\bigr)\nonumber\\
	&&  + \tfrac{1}{1152} d_4^4 \bigl( \csch(\gamma)\bigr)^7 \bigl( \sech(\gamma)\bigr)^7 \bigl(\mathcal{P} \cosh(\gamma) + P_1 \sinh(\gamma)\bigr)^2 \bigl(31504 P_1^3 \mathcal{P} - 3832 P_1 \mathcal{P}^3 \nonumber\\
	&& - 8 P_1 \mathcal{P} (5479 P_1^2 + 703 \mathcal{P}^2) \cosh(2 \gamma) + 16 P_1 \mathcal{P} (911 P_1^2 - 125 \mathcal{P}^2) \cosh(4 \gamma)\nonumber\\
	&&  - 8 P_1 \mathcal{P} (281 P_1^2 + 17 \mathcal{P}^2) \cosh(6 \gamma)  + 72 P_1 \mathcal{P}^3 \cosh(8 \gamma) \nonumber\\
	&&+ 2 (-2227 P_1^4 - 306 P_1^2 \mathcal{P}^2 + 269 \mathcal{P}^4) \sinh(2 \gamma) + 2 (1769 P_1^4 + 4266 P_1^2 \mathcal{P}^2 + 253 \mathcal{P}^4) \sinh(4 \gamma)\nonumber\\
	&&  + 2 (-431 P_1^4 - 858 P_1^2 \mathcal{P}^2 + 97 \mathcal{P}^4) \sinh(6 \gamma) + 9 (- P_1^4 + 6 P_1^2 \mathcal{P}^2 + 3 \mathcal{P}^4) \sinh(8 \gamma)\bigr)\nonumber\\
	&&  -  \tfrac{1}{36} d_7^2 \bigl(\mathcal{P}  \csch(\gamma) + P_1  \sech(\gamma)\bigr)^4 \bigl(206 P_1 \mathcal{P} - 108 P_1 \mathcal{P} \cosh(2 \gamma) - 26 P_1 \mathcal{P} \cosh(4 \gamma)\nonumber\\
	&&  + 2 (-59 P_1^2 + 5 \mathcal{P}^2) \sinh(2 \gamma) - 13 (P_1^2 + \mathcal{P}^2) \sinh(4 \gamma)\bigr) \tanh(\gamma)\nonumber\\
	&&  -  \tfrac{1}{16} d_{11} d_4 \bigl(\mathcal{P}  \csch(\gamma) + P_1  \sech(\gamma)\bigr)^4 \bigl(182 P_1 \mathcal{P} - 96 P_1 \mathcal{P} \cosh(2 \gamma)\nonumber\\
	&&  - 22 P_1 \mathcal{P} \cosh(4 \gamma) + 2 (-53 P_1^2 + 5 \mathcal{P}^2) \sinh(2 \gamma) - 11 (P_1^2 + \mathcal{P}^2) \sinh(4 \gamma)\bigr) \tanh(\gamma)\Bigr),
\end{eqnarray}
and 
\begin{eqnarray}\label{TaaTbbfa}
	&&T^{\mu}{}_{ \mu } T^{\nu}{}_{ \nu } =\frac{d_4^2 \lambda^2 \bigl(P_1 + \mathcal{P}  \coth(\gamma)\bigr)^4}{\bigl( \coth(\gamma)\bigr)^2}  \\
	&& + \lambda^3 \Bigl(\frac{8 d_4^3 \bigl(-2 P_1 + \mathcal{P}  \coth(\gamma)\bigr) \bigl(P_1 + \mathcal{P}  \coth(\gamma)\bigr)^4 \bigl( \sech(\gamma)\bigr)^3}{3  \coth(\gamma)}+ \frac{8 d_4 d_7 \bigl(P_1 + \mathcal{P}  \coth(\gamma)\bigr)^5}{3 \bigl( \coth(\gamma)\bigr)^3}\Bigr)\nonumber\\
	&&  + \lambda^4 \Bigl(\tfrac{1}{36} d_4^4 \bigl( \csch(\gamma)\bigr)^6 \bigl( \sech(\gamma)\bigr)^6 \bigl(\mathcal{P} \cosh(\gamma) + P_1 \sinh(\gamma)\bigr)^4  \nonumber\\
	&&\times \bigl(-479 P_1^2 + 113 \mathcal{P}^2 + 140 (4 P_1^2 + \mathcal{P}^2) \cosh(2 \gamma)  \nonumber\\
	&&+ 27 (-3 P_1^2 + \mathcal{P}^2) \cosh(4 \gamma) - 236 P_1 \mathcal{P} \sinh(2 \gamma) - 54 P_1 \mathcal{P} \sinh(4 \gamma)\bigr)  \nonumber\\
	&& + \frac{2 d_4^2 d_7 \bigl(P_1 + \mathcal{P}  \coth(\gamma)\bigr)^5 \bigl(-113 P_1 + 43 \mathcal{P}  \coth(\gamma)\bigr) \bigl( \sech(\gamma)\bigr)^3}{9 \bigl( \coth(\gamma)\bigr)^2}\nonumber\\
	&& + \frac{3 d_{11} d_4 \bigl(P_1 + \mathcal{P}  \coth(\gamma)\bigr)^6}{\bigl( \coth(\gamma)\bigr)^4}  + \frac{16 d_7^2 \bigl(P_1 + \mathcal{P}  \coth(\gamma)\bigr)^6}{9 \bigl( \coth(\gamma)\bigr)^4}\Bigr).\nonumber
\end{eqnarray}

	\if{}
	\bibliographystyle{abe}
	\bibliography{references}{}
	\fi
	
	\providecommand{\href}[2]{#2}\begingroup\raggedright\endgroup
\end{document}